\documentclass[usenatbib]{mn2e_letter}
\usepackage{amssymb, amsmath}
\usepackage{graphicx}

\newcommand{\dif}{\mathrm{d}}
\newcommand{\degree}{\ensuremath{^\circ}}
\def\la{\langle}
\def\ra{\rangle}

\def\Mo               {\hbox{$M_{\odot}$}}

\def\omc{\omega_{\rm cond}}
\def\omv{\omega_{\rm visc}}
\def\omb{\omega_{\rm buoy}}
\def\sigmat{\tilde \sigma}
\def\kperp{k_{\perp}}
\def\kpar{k_{\parallel}}

\def\kms{\textrm{km}\;\textrm{s}^{-1}}
\def\kpc{\textrm{kpc}}
\def\Pr{\textrm{Pr}}
\def\kevcm{\textrm{keV}\,\textrm{cm}^2}
\def\bhat{\boldsymbol{\hat{b}}}
\def\khat{\boldsymbol{\hat{k}}}
\def\vhat{\boldsymbol{\hat{v}}}

\newcommand       \be           {\begin{equation}}
\newcommand       \ee           {\end{equation}}
\def\onehalf{\frac{1}{2}}

\newcommand{\acknowledgments}{\begin{small}
    \section*{Acknowledgments}\end{small}}
\newcommand\altaffilmark[1]{$^{#1}$}
\newcommand\altaffiltext[1]{$^{#1}$}

\begin{document}
\title[Anisotropic Viscosity in the Intracluster Medium]{The Effects of Anisotropic Viscosity on Turbulence and Heat Transport in the Intracluster Medium}
\author[Parrish, McCourt, Quataert, and Sharma]{
\parbox[t]{\textwidth}{
Ian J. Parrish\altaffilmark{1}\thanks{E-mail: iparrish@astro.berkeley.edu}, Michael McCourt, Eliot Quataert, and Prateek Sharma\altaffilmark{2}}
\vspace*{6pt}\\
\altaffiltext{1}{Department of Astronomy and Theoretical Astrophysics
  Center, University of California
  Berkeley, Berkeley, CA 94720} \\
\altaffiltext{2}{Present Address: Department of Physics, Indian Institute of Science, Bangalore 560012, India} \\
}
%\affiliation{Astronomy Department \& Theoretical Astrophysics Center, 601 Campbell Hall, University of California Berkeley, CA 94720, USA; iparrish@astro.berkeley.edu}
%\altaffiltext{1}}
\maketitle
\begin{abstract}

  In the intracluster medium (ICM) of galaxy clusters, heat and
  momentum are transported almost entirely along (but not across)
  magnetic field lines.  We perform the first fully self-consistent
  Braginskii-MHD simulations of galaxy clusters including both of
  these effects.  Specifically, we perform local and global
  simulations of the magnetothermal instability (MTI) and the
  heat-flux-driven buoyancy instability (HBI) and assess the effects
  of viscosity on their saturation and astrophysical implications.  We
  find that viscosity has only a modest effect on the  saturation of the MTI. As in previous calculations, we find
  that the MTI can generate nearly sonic turbulent velocities in the
  outer parts of galaxy clusters, although viscosity somewhat suppresses the magnetic field amplification.
  At smaller radii in cool-core clusters, viscosity can
  decrease the linear growth rates of the HBI.  However, it has less
  of an effect on the HBI's nonlinear saturation, in part because
  three-dimensional interchange motions (magnetic flux tubes slipping
  past each other) are not damped by anisotropic viscosity.  In global
  simulations of cool core clusters, we show that the HBI robustly
  inhibits radial thermal conduction and thus precipitates a cooling
  catastrophe. The effects of viscosity are, however, more important
  for higher entropy clusters.  We argue that viscosity can contribute
  to the global transition of cluster cores from cool-core to non
  cool-core states: additional sources of intracluster turbulence,
  such as can be produced by AGN feedback or galactic wakes, suppress
  the HBI, heating the cluster core by thermal conduction; this makes
  the ICM more viscous, which slows the growth of the HBI, allowing
  further conductive heating of the cluster core and a transition to a
  non cool-core state.
\end{abstract}
\section{Introduction}\label{sec:intro}
 Clusters of galaxies are the largest gravitationally bound objects in the universe, and as such, they potentially provide sensitive tests of cosmological parameters.  They are filled with a hot, dilute, magnetized plasma, the intracluster medium (ICM), that emits copious X-rays.   Observations of the ICM provide an interesting and unique window on problems ranging from constraining dark energy to understanding the accretion and feedback processes for supermassive black holes.  
%Many of these measurements require inferring the total mass of a galaxy cluster from the $\sim \!12$--13\% of the total mass that resides in the ICM.  In order to reliably study all of these processes, we must have a detailed and accurate understanding of the ICM.
 
 The plasma in the ICM has temperatures ranging from 1--15 keV and
 number densities from $10^{-4}$ to $10^{-1}\;\textrm{cm}^{-3}$.  This
 dilute plasma has a magnetic field that has been estimated to range
 from 0.1--10 $\mu$G \citep{ct02}.  With these parameters the mean
 free path of electrons along the magnetic field line is
 $\gtrsim10^{12}$ times larger than the gyroradius at all radii in the
 ICM.  Ions have a similar separation of scales.  The mean free path
 is, however, always shorter than the local scale height; therefore, a
 fluid description of the plasma (as opposed to a collisionless
 description) is appropriate and the ICM can be described by the
 Braginskii-MHD equations \citep{brag65}.  These equations are the
 standard ideal MHD equations supplemented with anisotropic conduction
 due to the electron heat flux and anisotropic momentum transport due
 to the ion viscosity along the magnetic field.  In the ICM, transport
 perpendicular to the local magnetic field is negligible.
 
 As a result of the anisotropic heat transport in the ICM, the
 Schwarzschild criterion for convective instability---that the
 entropy increase in the direction of gravity---is replaced by a
 criterion on temperature.  In recent years, two buoyant instabilities
 have been discovered that drive convection with the temperature
 gradient as a source of free energy.  The first instability, the
 magnetothermal instability (MTI), was described in \citet{bal00} and
 has been simulated in two and three dimensions \citep{ps05, ps07b,
   mpsq11}.  The MTI is unstable when the temperature gradient and
 gravity are in the same direction and grows fastest for a magnetic
 field perpendicular to gravity.  The MTI operates in the outskirts of
 galaxy clusters and has been found to drive vigorous convection that
 can provide over 30\% of the pressure support near the virial radius
 \citep{pmqs11}.  The second instability, the heat-flux-driven
 buoyancy instability (HBI) was described in \citet{quat08} and has
 been simulated in local simulations in 2D and 3D \citep{pq08}.  The
 HBI is unstable when the temperature gradient and gravity point in
 opposite directions and has the fastest growth for a magnetic field
 parallel to gravity.  The HBI operates in the centers of cool-core
 clusters and saturates by reorienting the magnetic field to be
 perpendicular to gravity, greatly reducing the effective radial
 conductivity and hastening a cooling catastrophe \citep{pqs09,
   bog09}.  The interaction of the HBI with turbulence can help explain the bimodality observed between cool core
 and non-cool core clusters \citep{pqs10, rus10}. 
% We revisit all of
% these topics in this work, with a particular focus on the effect of
% anisotropic viscosity.

 None of these previous numerical studies have self-consistently
 included viscosity because the ratio of viscous to thermal diffusion,
 the Prandtl number,\footnote{This quantity should not be
  confused with the often-discussed magnetic Prandtl number in which
  the thermal diffusivity is replaced by the electrical resistivity.
  The magnetic Prandtl number is very large in the ICM.} for a hydrogenic plasma is
\begin{equation}
\textrm{Pr} = \frac{\nu}{\chi}\approx 0.01.
\label{eqn:Pr}
\end{equation}
where $\nu$ and $\chi$ are the diffusion coefficients for momentum (by ions only) and
thermal energy (by electrons only), respectively.  Pr $\sim 0.01$ is relatively small, and
thus the effects of viscosity were expected to be small in comparison
with those of conduction.\footnote{The effective value of the Prandtl number that enters the perturbed total energy equation in the linear analysis corresponds to $\textrm{Pr}_{\textrm{eff}}=0.02$.  This is because $\mu = 0.5$:  only electrons participate in conduction, while both electrons and ions contribute to the total thermal energy.  The net effect is that $\chi$ is smaller for the MHD fluid by a factor of 2.  In our calculations we take $\textrm{Pr} = 0.01$ and $\mu = 0.5$.} However, the
Reynolds number is fairly small in the ICM:
\begin{equation}
\textrm{Re} = 4\left(\frac{U}{100\;\kms}\right) \left(\frac{L}{100\;\kpc}\right) \left(\frac{n_e}{0.05\;\textrm{cm}^{-3}}\right)
\left(\frac{k_B T}{3\;\textrm{keV}}\right)^{-5/2},
\label{eqn:Reynolds}
\end{equation}
and thus it is not so clear that the effects of viscosity can be
neglected.  \citet[][]{kunz11} (hereafter K11) recently extended the
linear dispersion relation for the MTI and HBI to include anisotropic
viscosity and provides an intuitive, physical explication of its
effects.  The most important is that the growth rates of the HBI can
be suppressed by viscosity in the limit of very rapid conduction (and
thus very rapid viscous damping).  Isotropic viscosity has been utilized in
a small number of previous numerical studies; e.g., \citet{reynolds05}
studied the effect of viscosity on the shapes of rising AGN-blown
bubbles. More recently, \citet{dongstone09} showed that it is critical
to consider anisotropic viscosity, rather than isotropic viscosity, in
such calculations because the former is much less effective at
suppressing the Rayleigh-Taylor instability.

In this paper, we present fully self-consistent 2D and 3D
Braginskii-MHD simulations of the ICM, focusing on the evolution of
the MTI and HBI.  We introduce our computational methods in
\S\ref{sec:physics}.  In \S\ref{sec:local} we present local 2D and 3D
simulations of the HBI and MTI to provide physical insight into the
role of viscosity.  We then use global calculations to study the effect of viscosity on the
MTI in the outskirts of galaxy clusters and the role of the viscous
HBI in cluster cores in \S5 \& \S6, respectively. In the appendix we
describe our numerical method for anisotropic viscous transport and
discuss the numerical verification of this algorithm.
\section{Method and Models}\label{sec:methods}
We solve the usual equations of magnetohydrodynamics (MHD) with the
addition of anisotropic thermal conduction and anisotropic viscous transport. The MHD equations in
conservative form are
\begin{equation}
\frac{\partial \rho}{\partial t} + \boldsymbol{\nabla}\cdot\left(\rho \boldsymbol{ v}\right) = 0,
\label{eqn:MHD_continuity}
\end{equation}
\begin{equation}
\frac{\partial(\rho\boldsymbol{v})}{\partial t} + \boldsymbol{\nabla}\cdot\left[\rho\boldsymbol{vv}+\left(p+\frac{B^{2}}{8\pi}\right)\mathbf{I} -\frac{\boldsymbol{BB}}{4\pi} +\mathsf{\Pi} \right] + \rho\boldsymbol{g}=0,
\label{eqn:MHD_momentum}
\end{equation}
\begin{eqnarray}
\label{eqn:MHD_energy}
\frac{\partial E}{\partial t} &+& \boldsymbol{\nabla}\cdot\left[\boldsymbol{v}\left(E+p+\frac{B^{2}}{8\pi}\right) - \frac{\boldsymbol{B}\left(\boldsymbol{B}\cdot\boldsymbol{v}\right)}{4\pi} + \mathsf{\Pi}\cdot\boldsymbol{v} \right] \\ \nonumber
&+&\boldsymbol{\nabla}\cdot\boldsymbol{Q} +\rho\boldsymbol{\nabla}\Phi\cdot\boldsymbol{v}
=- \mathcal{L},
\end{eqnarray}
\begin{equation}
\frac{\partial\boldsymbol{B}}{\partial t} - \boldsymbol{\nabla}\times\left(\boldsymbol{v}\times\boldsymbol{B}\right)=0,
\label{eqn:MHD_induction}
\end{equation}
which are the equations of conservation of mass, momentum, and energy and the induction equation, respectively.  The total energy $E$ is given by
\begin{equation}
E=\epsilon+\rho\frac{\boldsymbol{v}\cdot\boldsymbol{v}}{2} + \frac{\boldsymbol{B}\cdot\boldsymbol{B}}{8\pi},
\label{eqn:MHD_Edef}
\end{equation}
where $\epsilon=p/(\gamma-1)$.  Throughout this paper, we assume
$\gamma=5/3$.  The anisotropic electron heat flux is given by
\begin{equation}
\boldsymbol{Q} = - \kappa_{\textrm{Sp}} \boldsymbol{\hat{b}\hat{b}}\cdot\boldsymbol{\nabla}T,
\label{eqn:coulombic}
\end{equation}
where $\kappa_{\textrm{Sp}}$ is the Spitzer conductivity \citep{spitz62} and
$\boldsymbol{\hat{b}}$ is a unit vector in the direction of the
magnetic field.  The Spitzer conductivity can be written as $\kappa_{\textrm{Sp}} = n_e k_B \chi$, where $\chi$ is the actual diffusion coefficient (in units of $L^2T^{-1}$).
The viscous stress tensor is given by
\begin{equation}
\mathsf{\Pi} \equiv -3\rho\nu\left[\bhat\bhat:\nabla \boldsymbol{v}
-\frac{\nabla\cdot\boldsymbol{v}}{3}\right]
\left[\bhat\bhat - \frac{\mathsf{I}}{3}\right],
\label{eqn:viscosity_tensor}
\end{equation}
where $\nu$ is the microphysical momentum diffusion coefficient, often termed the kinematic viscosity.  Both transport coefficients are functions of temperature proportional to $T^{5/2}$, where we presume $T_e = T_i$.  The microphysics fixes the ratio of the transport coefficients to be $\Pr = 0.01$ as given by Equation (\ref{eqn:Pr}).  

The energy equation also includes a cooling term, $\mathcal{L}$.  The cooling function we adopt is from \citet{tn01} with the functional form  
\begin{equation}
\mathcal{L} = n_e n_p \Lambda(T),
\label{eqn:cooling1}
\end{equation}
with units of erg cm$^{-3}$ s$^{-1}$.  The temperature dependence is a fit to cooling dominated by Bremsstrahlung above 1 keV and metal lines below 1 keV with 
\begin{equation}
\Lambda(T) = \left[C_1(k_B T)^{-1.7} + C_2(k_B T)^{0.5} + C_3\right] 10^{-22},
\label{eqn:cooling2}
\end{equation}
where $C_1 = 8.6\times 10^{-3}$, $C_2 = 5.8\times 10^{-2}$, and $C_3 = 6.3 \times 10^{-2}$, for a metallicity of $Z = 0.3 Z_{\odot}$, with units of $[C_i]=\textrm{erg}\, \textrm{cm}^3\textrm{s}^{-1} $.   We use a mean molecular weight of $\mu \sim 0.62$ which corresponds to a metallicity of approximately $1/3$ solar.  

For our simulations we use the Athena MHD code \citep{gs08, sg08} combined with the anisotropic conduction methods of \citet{ps05} and \citet{sh07}.  The anisotropic viscosity is implemented in a very similar manner to conduction (see the appendix). The heating, cooling, and anisotropic conduction and viscosity are operator split and sub-cycled with respect to the MHD timestep.  The cooling simulations are implemented with a temperature floor  of $T = 0.05$ keV, below which UV lines become important, and the cooling curve fit is no longer accurate.  This temperature floor prevents the cooling catastrophe from going to completion.  

This paper will cover a variety of initial conditions from local Cartesian boxes to global cluster models.  Our initial conditions will thus be described briefly in subsequent sections with appropriate references for more details.  In each case, we have carried out a least one resolution study to assure numerical convergence.  Any deviations will be noted.  
\vspace{-0.7cm}
\section{Physics of the MTI and HBI with Viscosity} \label{sec:physics}
\begin{figure*}
\centering
\includegraphics[clip=true, scale=1.0]{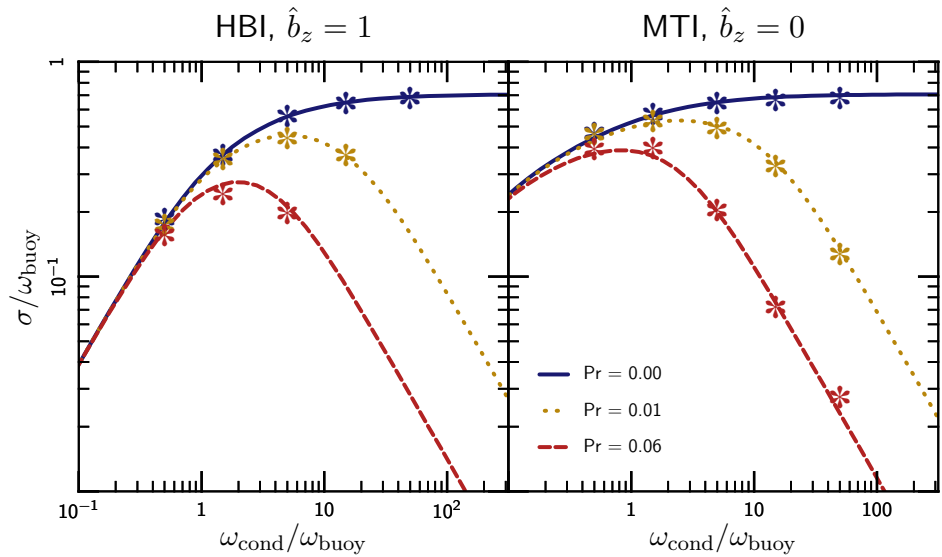}
\caption{The lines show the theoretical linear growth rates of the MTI
  (right) and HBI (left) for a variety of conduction frequencies and
  several Prandtl numbers for $\boldsymbol{k}$ at 45$\degree$ relative
  to the magnetic field.  This geometry is just an example and does
  not show the fastest growing mode for either the MTI or HBI.  In
  particular, for the MTI there are modes with growth rates comparable
  to the inviscid (Pr = 0) case even for $\omc \gg \omb$ (K11); this
  is not true for the HBI.  Real astrophysical plasmas have Pr = 0.01.
  Pr = 0.06 is shown only as a test of our numerical methods.  Low
  $\omc$ corresponds to larger scales at fixed conductivity.  Measured
  growth rates from 2D simulations are plotted as star symbols and
  agree well with the analytic results.} \label{fig:dr}
\end{figure*}
We begin with a qualitative description of the physics of the HBI and
MTI.  Despite the mathematical similarities of the instabilities, it
is useful to discuss each instability separately.  The MTI is most
unstable for horizontal ($\boldsymbol{B} \perp\boldsymbol{g}$)
magnetic fields with the temperature gradient $\dif T/\dif z < 0$.  An
upwardly displaced fluid element is connected by magnetic field lines
to a hotter region deeper in the atmosphere.  Heat flowing along the
field expands the fluid element relative to its surroundings, lowers
its density and buoyantly destabilizes the perturbation.  The upward
motion causes the magnetic field to be more aligned with the
background temperature gradient, leading to an instability.

The HBI, on the other hand, is most unstable for vertical ($\boldsymbol{B} \parallel\boldsymbol{g}$) magnetic
fields with the temperature gradient $\dif T/\dif z> 0$.  Imagine a
small displacement of fluid elements with a wavevector that has a
component both parallel and perpendicular to gravity (and
$\boldsymbol{B}$).  This configuration, illustrated in Figure 1 of
\citet{quat08}, has regions in which the magnetic field lines bunch
together and spread apart, leading to a converging and diverging heat
flux.  Rapid heat conduction along the perturbed magnetic field lines
causes an upwardly displaced fluid element to be heated (by tapping
into the background heat flux), leading to a buoyant runaway.
%Where the magnetic field lines converge, there is a net increase in
%heat flux, $+\delta \boldsymbol{Q}$, which again heats the region and
%drives an upward buoyant response. 
%This buoyant response causes the
%regions of converged field to be more converged, leading to a growing
%instability.

K11 carried out a full linear perturbation analysis on equations
\ref{eqn:MHD_continuity}--\ref{eqn:coulombic} and presents a
dispersion relation for the MTI and HBI including the effects of
anisotropic viscosity.  For clarity when comparing with our results,
we reproduce the K11 dispersion relation here in a slightly different
coordinate system.  We define gravity to be in the $z$-direction, the
initial magnetic field is in the $x$-$y$ plane and $\kperp^2 = k_x^2 +
k_y^2$ is the component of the wavevector perpendicular to
gravity. The dispersion relation for the growth rate, $\sigma$, is
given by
\begin{eqnarray}
\lefteqn{-\omv \left[ 1 - \left(\bhat \cdot \khat\right)^2 \right]} \\
&=& \frac{\sigmat^2 \left[ \sigmat^2\left(\sigma + \omc\right) + \sigma N^2\frac{\kperp^2}{k^2}  + \omc\omb^2 \mathcal{K} \right]}
{\sigma \left[ \sigmat^2\left(\sigma + \omc\right) + \left(\sigma N^2 + \omc\omb^2\right)
\frac{b_x^2k_y^2}{1-(\bhat \cdot \khat)^2}  \right]} ,\nonumber
\label{eqn:fulldispersion}
\end{eqnarray}
where $\sigmat^2 \equiv \sigma^2 - \left(\boldsymbol{k}\cdot\boldsymbol{v_A}\right)^2$, and $\boldsymbol{v_A} = \boldsymbol{B}/(4\pi \rho)^{1/2}$ is the Alfv\'{e}n velocity.  The Brunt-V\"ais\"al\"a oscillation frequency is given by
\begin{equation}
N^2 = -\frac{1}{\gamma \rho}\frac{\partial P}{\partial z}\frac{\partial \ln S}{\partial z},
\label{eqn:BV}
\end{equation}
where $S\equiv P \rho^{-\gamma}$, and corresponds to buoyant oscillations in an unmagnetized plasma (g-modes).  The characteristic frequency at which conduction and viscosity act are given by
\begin{eqnarray}
\omc &=& \frac{2}{5}\chi\left(\bhat \cdot \boldsymbol{k}\right)^2,\\
\omv &=& \frac{3}{2}\frac{v_{\textrm{th}}^2}{\nu_{ii}}\left(\bhat \cdot \boldsymbol{k}\right)^2,
\label{eqn:omc}
\end{eqnarray}
where $\chi$ is the thermal diffusivity and $\nu_{ii}$ is the ion collision frequency.  The buoyancy frequency is given by
\begin{equation}
\omb^2 = \left| g \frac{\partial \ln T}{\partial z} \right|,
\label{eqn:omb}
\end{equation}
which is roughly the fastest growth rate of the MTI or HBI.  Finally, the dimensionless geometric factor is given by
\begin{equation}
\mathcal{K} = \left(1 - 2b_z^2\right)\frac{\kperp^2}{k^2} + \frac{2b_x b_z k_x k_z}{k^2},
\label{eqn:K}
\end{equation}
where $b_x = B_x/B$ is the dimensionless magnetic field strength.  

We can greatly simplify this expression and improve our physical intuition by assuming that the magnetic field is weak (or, equivalently, that the perturbation wavelength in the cluster core $\lambda \gg v_{\mathrm{A}} / \tau \sim 2\;\textrm{kpc}$, where $\tau$ is a typical timescale for the instabilities to grow).   We also assume the perturbation lies in the plane containing $\boldsymbol{B}$ and $\boldsymbol{g}$, i.e. $k_y=0$.   In this limit, the dispersion relation simplifies to
\begin{eqnarray}
\lefteqn{\sigma\left(\sigma + \omc\right) \left[ \sigma + 
\omv \left(1- \left(\bhat \cdot \khat\right)^2  \right)\right]} \\
&+& \sigma N^2\frac{\kperp^2}{k^2} + \omc\omb^2\mathcal{K} = 0. \nonumber
\label{eqn:dr2}
\end{eqnarray}

Without viscosity, there is a fast conduction limit in which the
growth rate asymptotes to a constant $\sigma = \omb \kperp^2/k^2$;
with viscosity, however, no such limit exists.  In Figure \ref{fig:dr}
we plot the theoretical curves for the growth rates of both the HBI
and MTI as a function of the ratio $\omc/\omb$ for several Prandtl
numbers.  These curves assume a wave vector $\boldsymbol{k}$ at 45
degrees from the magnetic field $\boldsymbol{B}$; these represent
typical, not maximum, growth rates.  The ratio $\omc/\omb \propto k^2$
so that larger values of this ratio correspond physically to smaller
scales at fixed conductivity.  

Figure \ref{fig:dr} shows that the inviscid solution reaches an
asymptotic growth rate in the fast conduction limit; however, with
viscosity the growth rate slowly decreases as one moves to the
infinite conduction limit.  For the HBI, the fastest growth occurs on
scales satisfying $\omv \sim \omb$, i.e., $\omc \sim 6 \,\omb$, and
the maximum growth rate for $\textrm{Pr} = 0.01$ is about 70\% of the
asymptotic inviscid value.  This is independent of $\kperp$ unless
$\kperp \ll \kpar$.  For higher viscosities, the maximum growth rate
scales as $\sim \omb^2/\omv$.  The influence of viscosity on the MTI
is much more modest.  In particular, the fastest growing mode is
achievable even for $\omv,\omc \gg \omb$; this is not seen in Figure
\ref{fig:dr} because of the particular perturbation chosen.  One can, however, 
always find short wavelength MTI modes that grow at $\sim \omb$.

The microphysical plasma physics fixes the ratio of the diffusion
coefficients, the Prandtl number (eqn. [\ref{eqn:Pr}]), to be $\Pr =
0.01$.  This value corresponds to $\omv/\omc = 1/6$.  Because there
are MTI modes that grow rapidly even for $\omc \gg \omb$, we would a
priori not expect viscosity to have a significant effect on the
evolution of the MTI.  For the HBI, the viscous and inviscid growth
rates are similar only when $\omc/\omb\lesssim\;\textrm{10}$; these
larger scales in a physical system are only minimally affected by
viscosity.  We thus expect the HBI to be modified by viscosity only if
$\omc \gg 10 \;\omb$ on the largest scales of the system.  

\begin{figure*}
\centering
\includegraphics[clip=true, scale=0.9]{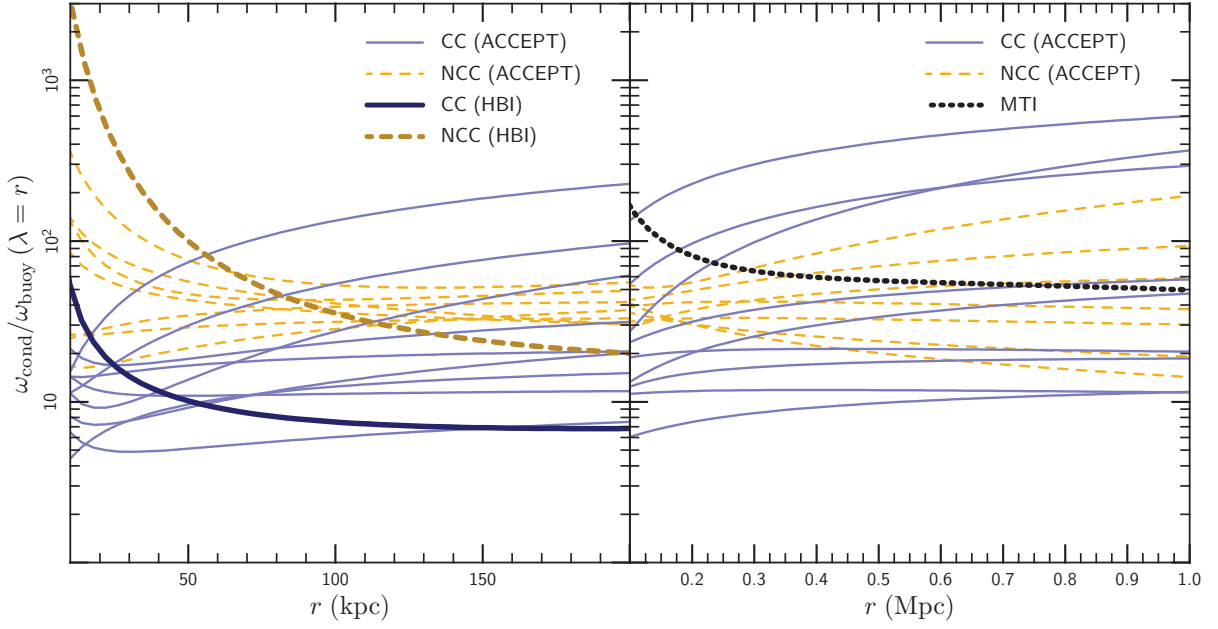}
\caption{\textit{Left:} The measured ratio of $\omc/\omb$ for the cores of clusters in the ACCEPT
  database, taking an average value of $|d \ln T/d \ln r| = 1/3$ in
  $\omb$ and $k = 2 \pi/r$ in $\omc$.  The latter choice corresponds
  to the conduction time across the local radius $r$ and represents
  the {\em smallest} value of $\omc/\omb$ at a given radius.
  Cool-core clusters have $\omc/\omb \sim 10$ in the bulk of the cool
  core.  Non-cool-core clusters have somewhat more rapid conduction
  and are thus more viscous as well.  We have also overplotted the
  values of $\omc/\omb$ in our global cluster models used for HBI (the
  CC and NCC HBI models).  \textit{Right:} The measured values of $\omc/\omb$ in the outskirts of clusters in the ACCEPT sample.  
  Our model cluster for the global MTI simulation is overplotted as well.   
%(valid for  $r>200$ kpc). 
Our fiducial model clusters are reasonably similar to observed
clusters.} \label{fig:tctb}
\end{figure*}

In Figure \ref{fig:tctb} we show the ratio of $\omc/\omb$ for observed
clusters using $k = 2 \pi/r$ as the conduction length scale and not
including any geometric factors, i.e. $\khat \cdot \bhat = 1$.  The
left panel focuses on the core of the cluster where the HBI operates
while the right panel is at larger radii where the MTI is present.
The values of $\omc/\omb$ in Figure \ref{fig:tctb} are characteristic
of the largest scales in the system.  Smaller scales are more
viscous/conducting, but are also where magnetic tension is most likely
to suppress the MTI/HBI.  Figure \ref{fig:tctb} also shows the ratio
$\omc/\omb$ (defined in the same way) for our model clusters used in
the global simulations in \S 5 \& 6.  The models are generally
representative of real clusters. 

The data in Figure \ref{fig:tctb} comes from the ACCEPT sample
\citep{cav09}; the analysis of the data follows the methodology
described in \citet{msqp11} with the clusters listed in Table 2 of
that work.   These examples span relatively massive NCC clusters all the way down  to groups.  Cool-core (CC) clusters have $\omc/\omb \sim 5$--30 in the
bulk of the core between a few--200 kpc.  By comparing the real CC
clusters to Figure \ref{fig:dr} we see that viscosity only modestly
reduces the growth rates of the HBI.  Non-cool-core (NCC) clusters are
more conducting/viscous as a result of the hotter and lower density
cores.  The effects of viscosity on the HBI are theoretically more likely to be
significant in NCC clusters.

The star symbols in in Figure \ref{fig:dr} show good agreement between the measured growth
rates in our simulations and the analytic results; this represents a strong
test of our algorithms for anisotropic conduction and viscosity.  We
discuss this in more detail in the appendix.
\section{Local Simulations}\label{sec:local}
\subsection{Initial Conditions}
\begin{figure*}
\centering
\includegraphics[clip=true, scale=1.0]{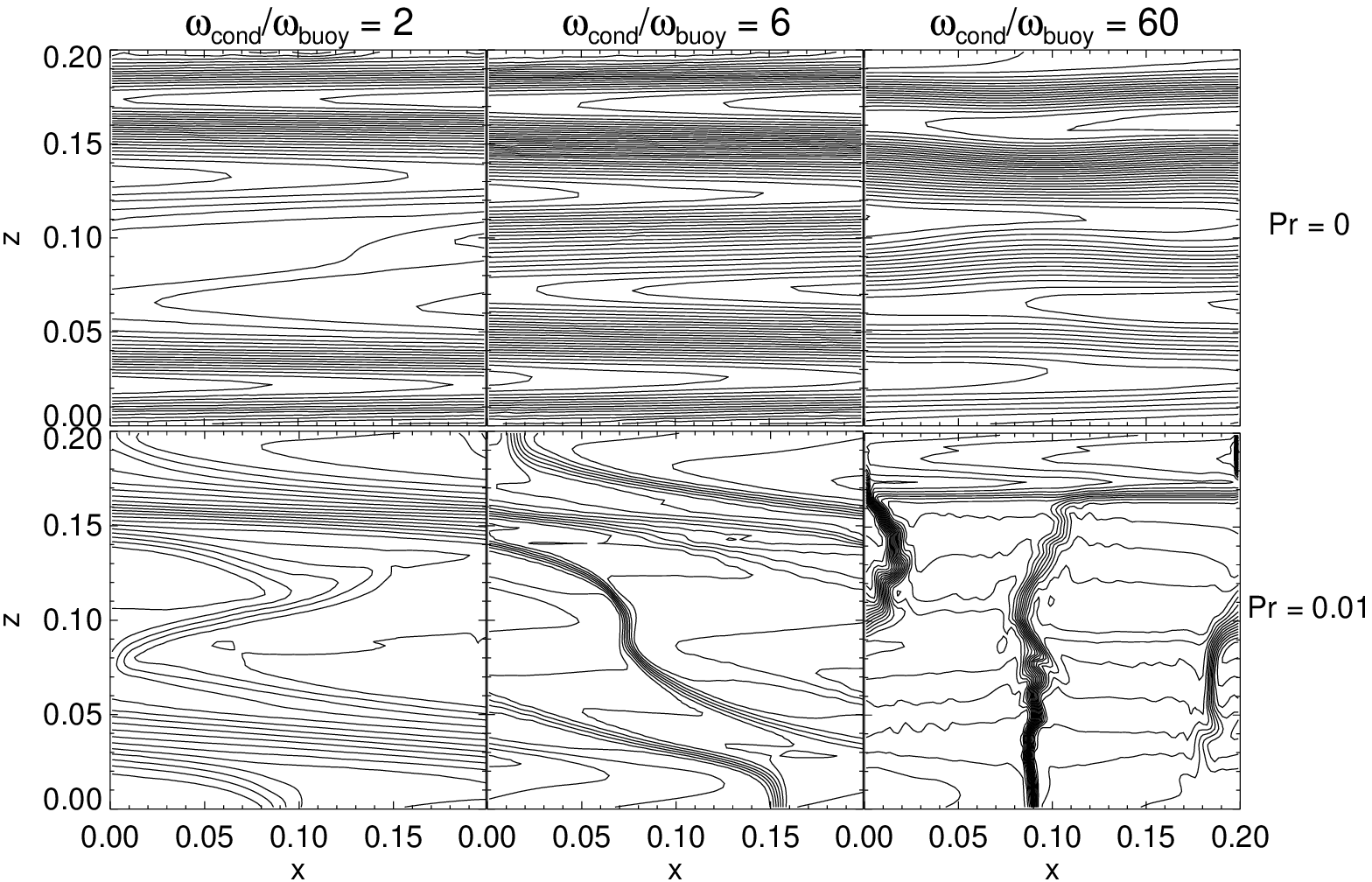}
\caption{Snapshots of the magnetic field lines from 2D HBI
  simulations.  The top row displays inviscid simulations, while the
  bottom row includes anisotropic viscosity.  The conductivity
  increases from left to right; the snapshots at a given conductivity
  are shown at the same absolute time for both the inviscid and
  viscous simulation (30 HBI growth times for the inviscid model). At
  low $\omc/\omb$, viscosity has little effect on the nonlinear
  evolution of the HBI; however, for $\omc/\omb \sim 60$, anisotropic
  viscosity causes the magnetic field to bunch into pronounced
  vertical structures.  This effect is much weaker in 3D simulations
  relative to the 2D results shown here (see
  Fig. \ref{fig:hbi2d3dcomp}).} \label{fig:hbi2d}
\end{figure*}
It is illustrative to start with the simplest possible experiments,
namely local two- and three-dimensional boxes with system sizes $L
\lesssim$ the scale-height $H$.  For this section we work in units
with $k_B=m_p=1$, $g_0 = -1$, and with a hydrogenic plasma that has
$\mu = 1/2$.  Our initial conditions are fully detailed in
\citet{mpsq11}.  For the HBI, we start with a simple hydrostatic
equilibrium that is linearly unstable:
\begin{eqnarray}
T(z) &=& T_0(1+z/H),\\
\rho(z) &=& \rho_0 (1+z/H)^{-3},
\label{eqn:HBI_IC}
\end{eqnarray}
where we set $T_0=\rho_0 = 1$ and choose a scale height of $H=2$.  Our boxes are of size $L= 0.2$ in each dimension with a resolution of $(96)^2$ or $(96)^3$.  Results at this resolution are very well converged by all metrics.  The vertical boundary conditions fix the temperature to its initial value, and the horizontal boundary conditions are periodic.   The initial magnetic field is weak and vertical, $B_0 = 10^{-6} \hat{\boldsymbol{z}}$.

For the local MTI simulations we utilize the set-up from \S3.3 of \citet{mpsq11} given by
\begin{eqnarray}
T(z) &=& T_0 \exp\left[\frac{S\omb^2}{g_0}\left(1- \mathrm{e}^{z/S}\right)     \right],\\
g(z) &=& g_0 \mathrm{e}^{-z/S},
\label{eqn:MTI_IC}
\end{eqnarray}
with $T_0 = g_0 = 1$ and  $S=3$.  We select $\omb^2 =1/2$ and solve numerically for $\rho(z)$ to ensure hydrostatic equilibrium.   These boxes are of size $H/2$ with an initially weak horizontal magnetic field, $B_0 = 10^{-6} \hat{\boldsymbol{z}}$.  Convergence is a bit more subtle in these  simulations, but $128^3$ is reasonably well-converged \citep{mpsq11}.  

For the HBI, the atmosphere satisfies the Schwarzschild criterion
($\dif S/\dif z > 0$) and would be buoyantly stable in the absence of
anisotropic conduction.  In these calculations we fix the
diffusivities so that the conduction frequency across the box is a
chosen constant relative to the buoyancy frequency, and show results
for different values of $\omc/\omb$ for both the inviscid case and the
physical case with $\Pr = 0.01$.

\vspace{-0.4cm}
\subsection{Nonlinear Saturation}
\subsubsection{HBI}
We begin by examining the nonlinear saturation of the HBI with 2D
single mode simulations.  The magnetic field lines from these
simulations are visualized in Figure \ref{fig:hbi2d}.  Each vertical
column compares inviscid and viscous simulations at the time $t= 30
\;t_{\textrm{HBI}}$.  This is well into the non-linear regime, in
which the statistical properties of the plasma do not evolve strongly
with time.  We are able to simulate a variety of physical scales in
these local simulations by changing the constant conductivity.  We
then label the simulation with a ratio of $\omc/\omb$ calculated using
the box size, $\lambda=L$, for the conduction length in $\omc$.  For
$\omc/\omb=2$, the leftmost column shows that viscosity does not
strongly change the saturated state.  The inviscid cases shown in the
top row all show qualitatively no difference in the nonlinear
saturated state as a function of $\omc/\omb$; however, the effect of anisotropic viscosity becomes
clear for the viscous case with $\omc/\omb=60$ (\textit{bottom
  right}).   This case shows
prominent, highly-bunched vertical magnetic field structures that are
never seen in the inviscid case.  This bunching occurs because the
field lines cannot slip past each other in 2D, a point we will return
to shortly.
\begin{figure*}
\centering
\includegraphics[clip=true, scale=0.45, angle=-90]{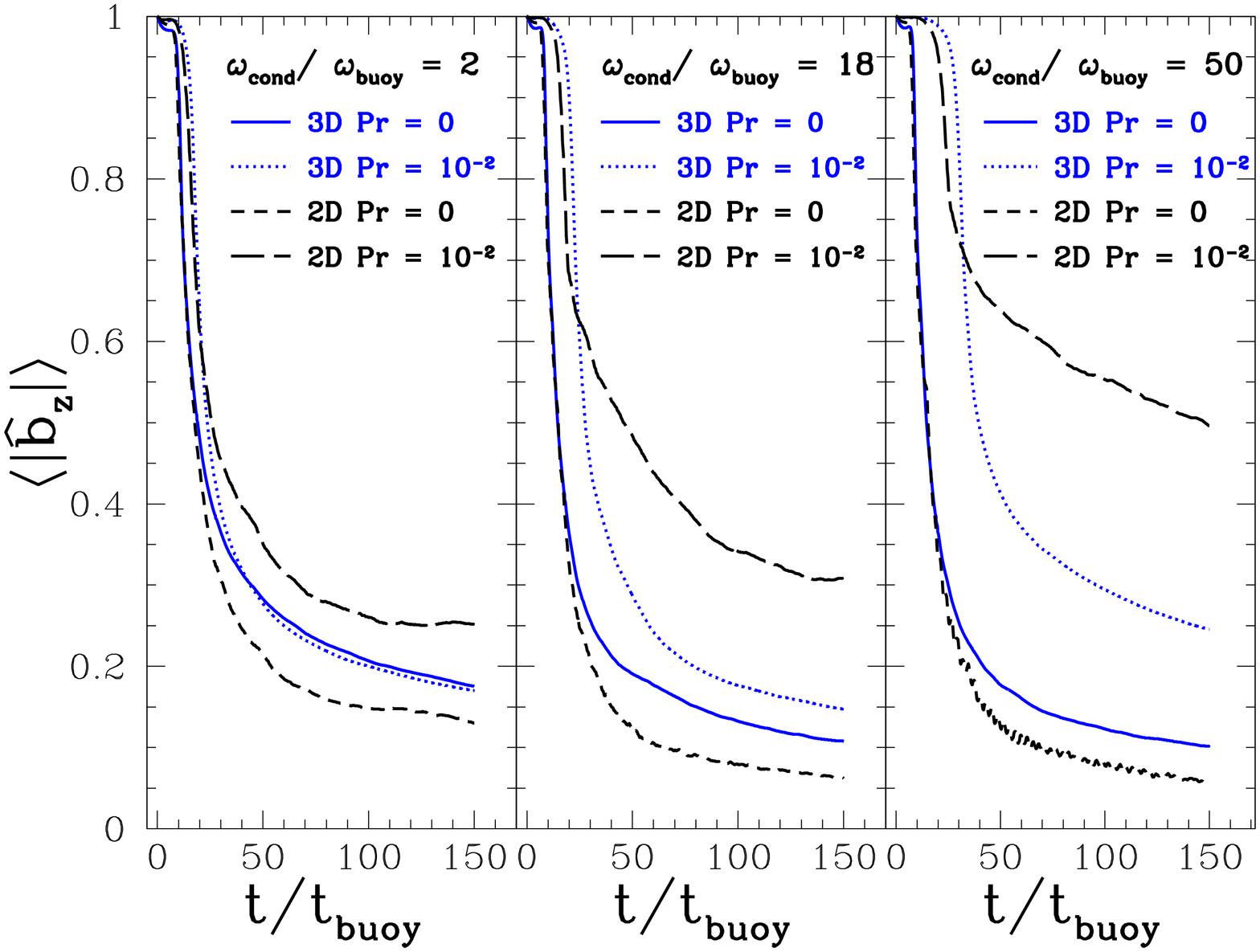}
\caption{The nonlinear evolution of the magnetic field due to the HBI
  for $\omc/\omb = 2$, 18, and 50 (\textit{left} to \textit{right}),
  in 2D and 3D local simulations ($L = 0.1H$).  We
  quantify the field evolution using the vertical component of the
  magnetic field.  The differences between the viscous and inviscid
  simulations are much smaller in 3D (\textit{blue lines}) than in 2D
  (\textit{black lines}).  We attribute this to the presence of
  interchange modes in 3D. Even for $\omc/\omb \sim 50$ the nonlinear
  evolution of the HBI in 3D is similar to that in inviscid
  simulations, with the primary difference being that the nonlinear
  satuation is delayed because of the reduced linear growth
  rate.} \label{fig:hbi2d3dcomp}
\end{figure*}

\begin{figure}
\centering
\includegraphics[clip=true, scale=0.45]{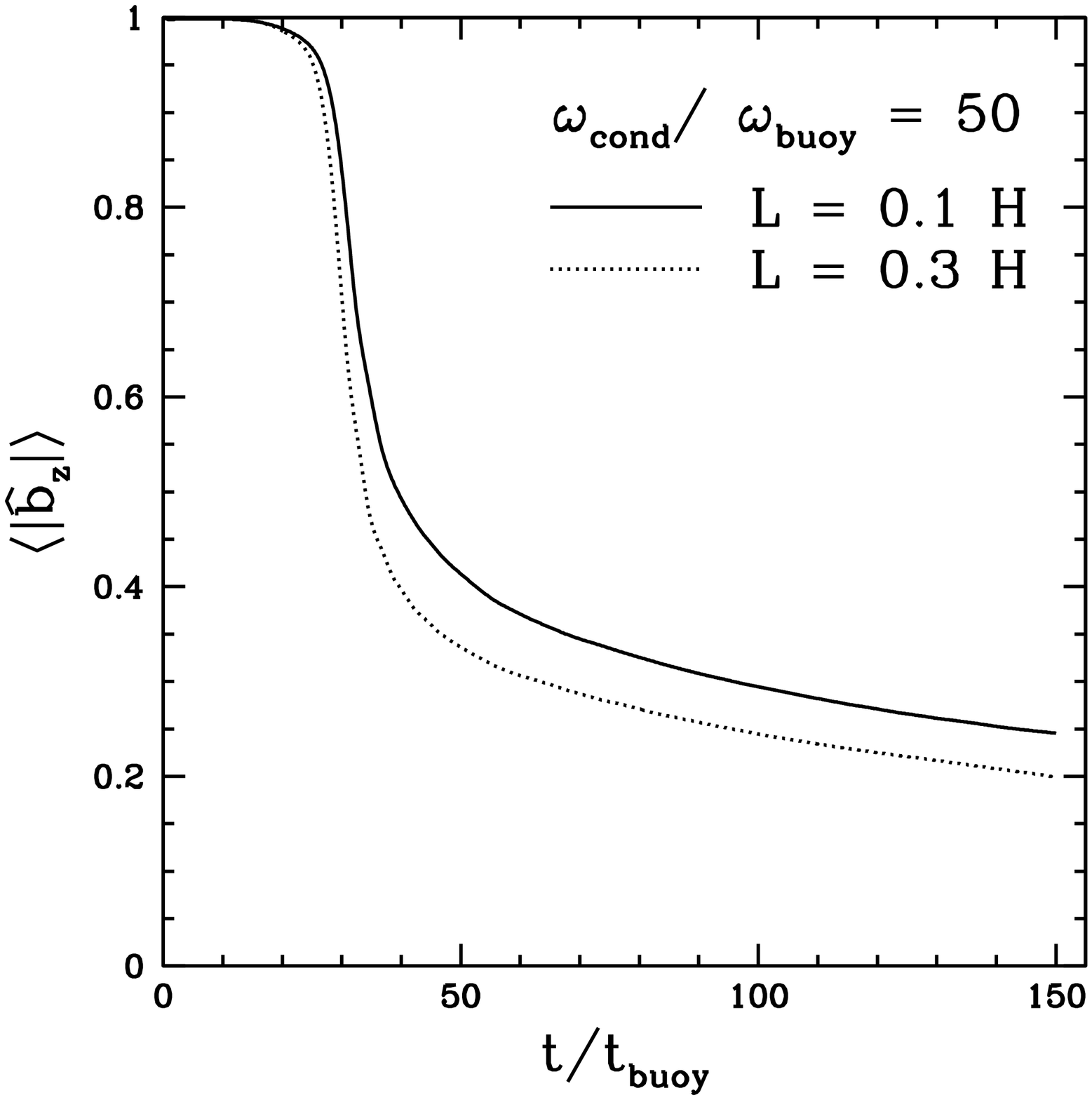}
\caption{The evolution of the magnetic field orientation for the viscous HBI ($\textrm{Pr} = 0.01$) for two different domain sizes relative to the temperature scale height ($L/H$).  To keep the ratio of $\omc(k=2\pi/L)/\omb = 50$ fixed, the simulation with $L/H = 0.3$ (\textit{dotted line}) has a conductivity and viscosity that are 9 times larger than the simulation with $L/H=0.1$ (\textit{solid line}).   For local simulations ($\L \lesssim H$) the linear and nonlinear evolution are nearly independent of box size for fixed $\omc/\omb$.} \label{fig:hbi-boxsize}
\end{figure}

In order to better understand the limitations of 2D simulations, it is
useful to compare the 2D and 3D evolution for identical parameters.
Figure \ref{fig:hbi2d3dcomp} shows the exact same simulations
performed in 2D and 3D for three different conductivities. For the
HBI, the results for simulations with different box sizes are
essentially the same provided that the same value of
$\omc(k=2\pi/L)/\omb$ is used \citep{mpsq11}.  We demonstrate this result explicitly in Figure \ref{fig:hbi-boxsize} which compares the magnetic geometry evolution for the viscous HBI in two different box sizes.  To keep the ratio of $\omc/\omb = 50$ fixed, the simulation with $L/H = 0.3$ has a conductivity and viscosity that are 9 times larger than the simulation with $L/H=0.1$.  The evolution, especially at linear and early nonlinear times, is nearly indistinguishable.  At very late times, the larger box has experienced slightly more magnetic field evolution as a result of the different modes in the domain relative to the scale height.  We conclude from Figure \ref{fig:hbi-boxsize} that local simulations of the HBI with $L \lesssim H$ are independent of domain size; the results depend instead on the value of $\omc/\omb$, the fundamental dimensionless parameter that quantifies the influence of conduction and viscosity.   Thus although the boxes
in Figure \ref{fig:hbi2d3dcomp} are only $\sim 0.1\, H$ in size, the
local simulations with $\omc/\omb = 18$ \& 50 are quite indicative of how
the HBI would evolve on large scales in the ICM (see Fig. \ref{fig:tctb}). 
The large scales are the most important because both tension and viscosity suppress the growth of the HBI on smaller scales.  We justify this interpretation of the local HBI simulations using global cluster models in \S\ref{hbicluster}.

Figure \ref{fig:hbi2d3dcomp} shows the mean magnetic field direction,
$\la |\bhat_z |\ra$, which is related to the effective vertical
thermal conductivity as $f_{\textrm{Sp}}\sim \la \bhat_z^2\ra$.  In 2D
the ability of the HBI to reorient that magnetic field is retarded by
viscosity relative to the inviscid case, especially for $\omc/\omb =
50$ (\textit{right}).  However, in 3D the effect of the viscosity is
much less pronounced.  The biggest effect of viscosity is that for
$\omc/\omb = 50$, the initial growth of the HBI is slower; however,
the instability still produces a significant rearrangement of the
magnetic field by $t \sim \, 50\, t_{\rm buoy}$.   We note that our initial velocity perturbations have white noise amplitudes with a mean magnitude of $5\times 10^{-4} c_s$.  For a larger and more realistic perturbation, the linear phase would complete more quickly.   

What is the cause of the striking difference in the 2D and 3D
evolution of the HBI with viscosity?  Recall in Figure \ref{fig:hbi2d}
that the magnetic field lines strongly bunched up in 2D and could not
slip past each other.  This type of motion is known as the interchange
mode and can be thought of as two magnetic flux tubes slipping past
each other.  Interchange motions have $\boldsymbol{v} \perp
\boldsymbol{B}$ and therefore are not damped at all by anisotropic
viscosity, although they would be damped by an isotropic viscosity.
In 3D the magnetic field lines are able to slip past each other, and
the HBI is able to continue to reorient the magnetic field.  This
effect is seen in a different guise in studies of the magnetic
Rayleigh-Taylor (RT) instability \citep{stoneRT}.  In 2D, magnetic
tension suppresses the RT instability below a critical wavelength;
however, in 3D magnetic tension does nothing to suppress the
interchange motions of flux tubes.  Thus, the instability 
grows despite the apparent wavelength cut-off in the dispersion
relation.  This uniquely 3D effect of anisotropic viscosity makes 2D
studies of nonlinear saturation irrelevant, and greatly reduces the
effect of viscosity relative to what one might na\"ively expect from
the dispersion relation.  It is particularly striking for $\omc/\omb =
50$ (\textit{right} panel in Fig. \ref{fig:hbi2d3dcomp}) the
non-linear evolution in 3D is quite similar with and without
viscosity, although the viscous evolution is delayed by the
considerably longer linear growth time.  
\begin{figure}
\centering
\includegraphics[clip=true, scale=0.45]{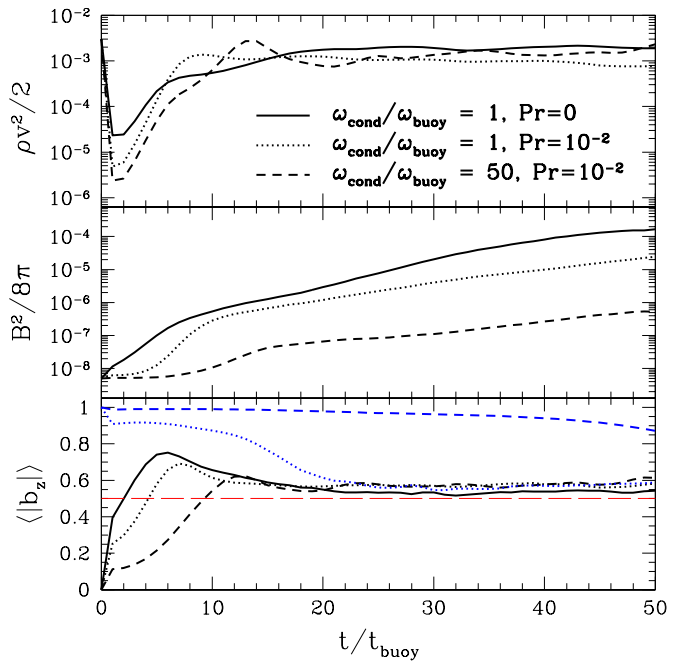}
\caption{The nonlinear evolution of the MTI for $\omc/\omb = 1$ and 50 in 3D ($L=H/2$) with and without viscosity.  We omit the $\omc/\omb=50$ inviscid simulation, as its evolution is almost exactly the same as the $\omc/\omb=1$ inviscid simulation.  The kinetic energy (\textit{top panel}) is largely independent of viscosity.  The magnetic energy generation (\textit{middle panel}) is suppressed for higher viscosities.  The magnetic geometry (\textit{bottom panel})  for initially horizontal magnetic fields (\textit{black lines}) becomes relatively isotropic for all the simulations, although slightly more vertical with higher viscosity.  Statistical isotropy corresponds to $\la | b_z | \ra = 0.5$ (\textit{red, long-dashed line}).  Simulations with initially vertical fields (\textit{blue lines}) are linearly stable but nonlinearly unstable \citep{mpsq11}; viscosity slows the reorientation of magnetic field in this case. } \label{fig:mtilocal}
\end{figure}
\subsubsection{MTI}
Next, we also examine the nonlinear evolution of the MTI
in 3D Cartesian simulations.  The dispersion relation shows that
the fastest growing modes of the MTI are less affected by viscosity
than they are for the HBI.  Figure \ref{fig:mtilocal} shows the results of our Cartesian simulations of the MTI with and without viscosity and reveals several interesting properties.  First, the saturated kinetic energies are largely independent of both the ratio of $\omc/\omb$ and whether anisotropic viscosity is included.   Note that the simulation with $\omc/\omb = 50$ is consistent with the range of clusters in the ACCEPT sample that we have examined (see the right panel of Figure \ref{fig:tctb}).  The fact that the kinetic energies are so similar even in the high-viscosity simulation, is likely the result of interchange-like turbulence.

Second, we find the very interesting result that the growth of the magnetic energy (middle panel of Figure \ref{fig:mtilocal}) is suppressed as the viscosity is increased.  This result can be understood in relatively easy terms: the increase in the magnetic field strength in this type of turbulence is proportional to the increase in the length of the magnetic field line, as it is stretched and tangled.  These parallel stretching motions are precisely those motions that are damped by anisotropic viscosity, thus suppressing the growth of the magnetic field.

Finally, we consider the reorientation of the magnetic geometry by the MTI (bottom panel of Figure \ref{fig:mtilocal}).  In the inviscid case, an initially horizontal magnetic field is re-oriented to be largely isotropic, $\la | b_z  | \ra = 1/2$ in 3D.  For modest viscosities $\omc/\omb =1$, there is little change in the saturated geometry; however, for higher conductivity and viscosities (e.g., $\omc/\omb = 50$) the magnetic field becomes slightly more vertical than isotropic (about a $10\%$ change in $\la |b_z | \ra$).  We can gain insight into this behavior by considering the evolution of simulations with an initially vertical field, a case that is linearly stable, but non-linearly unstable \citep[see Figure 6 of][]{mpsq11}.  An initial horizontal perturbation is purely Alfv\'enic in nature, and thus is unaffected by Braginskii viscosity.  A simulation with an initially vertical field and $\omc/\omb=1$ (blue dotted line) evolves towards an isotropic magnetic field configuration.  However, in doing so, the field develops a component parallel to the velocity, which is susceptible to viscous damping.  Thus, for an initially vertical magnetic field, the higher viscosity simulation ($\omc/\omb=50$, blue dashed line) evolves much more slowly than its less viscous counterpart.  It is precisely this partial stabilization of modes that initially have the field aligned with gravity that results in the radial bias seen in the high-conductivity, horizontal simulation.  
\section{Global Models of the MTI in Cluster Outskirts}\label{mticluster}
In this section we discuss the effect of Braginskii viscosity on the
evolution of the MTI in the outskirts of galaxy clusters using global cluster models.  Beyond
approximately the scale radius of a cluster, the temperature profile
of the ICM almost always declines with radius, thus making it unstable
to the MTI.  Recent work has shown that the MTI is able to drive large
turbulent velocities which provide non-thermal pressure support in
hot, massive clusters \citep{pmqs11}.  It is critical to understand
this non-thermal pressure support and its implications for measuring cluster
masses for use in cosmological parameter estimation through either
X-ray or Sunyaev-Zelovich (SZ) methods.

Our initial condition for this section is a spherically-symmetric, hot, massive cluster that resembles Abell 1576 with a mass of $1.6\times 10^{15}\;\Mo$.   We use a softened NFW profile with a scale radius of $r_s = 600$ kpc
and a softening radius of 70 kpc.  We initialize an atmosphere in
hydrostatic equilibrium using the entropy power law in the ACCEPT
database for Abell 1576: a central entropy $K_0 =
186\;\textrm{keV}\,\textrm{cm}^{2}$, $K_1 =
98\;\textrm{keV}\,\textrm{cm}^{2}$, and power-law exponent, $\alpha =
1.38$ \citep{cav09}.    If we
assume that our fiducial cluster is located at $z=0.1$, then for the
appropriate WMAP5 cosmological parameters $r_{500} =
1.09\;\textrm{Mpc}$, and the virial radius is $r_{200} = 1.6$ Mpc,
where $r_\Delta$ corresponds to an overdensity of $\Delta$ times the
critical density.  We do not include cooling, as we focus on the
portion of the cluster that is well outside the cooling radius.  For full simulation details, see \citet{pmqs11}.  By examining Figure \ref{fig:tctb}, we can see that the ratio of $\omc/\omb$ in our model is consistent with the range observed in real clusters in the ACCEPT database at the radii of interest ($r\gtrsim 200$ kpc). 

The simulations are carried out on a $(196)^3$ Cartesian grid in a
computational domain that extends from the center of the cluster out
to $\pm 1300$ kpc.  Within this Cartesian domain, we define a spherical
subvolume with a radius of 1225 kpc from which we extract
cluster properties.  In this
volume we initialize tangled magnetic fields with $\la|B|\ra =
10^{-8}\;\textrm{G}$ (plasma $\beta \sim 10^4$--$10^6$) and a Kolmogorov power spectrum. In order to
simulate clusters with a negative radial temperature gradient, we fix
the temperature at the peak of the cluster temperature profile
(approximately 10 keV at 200 kpc) and at the maximum radius (approximately 6.5 keV at 1225 kpc) of our model cluster
to the initially-computed temperature values.  Thus, we are imposing
Dirichlet boundary conditions on the temperature profile.  
This fixed temperature gradient then drives the continued evolution of
the MTI. 
\begin{figure}
\centering
\includegraphics[clip=true, scale=0.45]{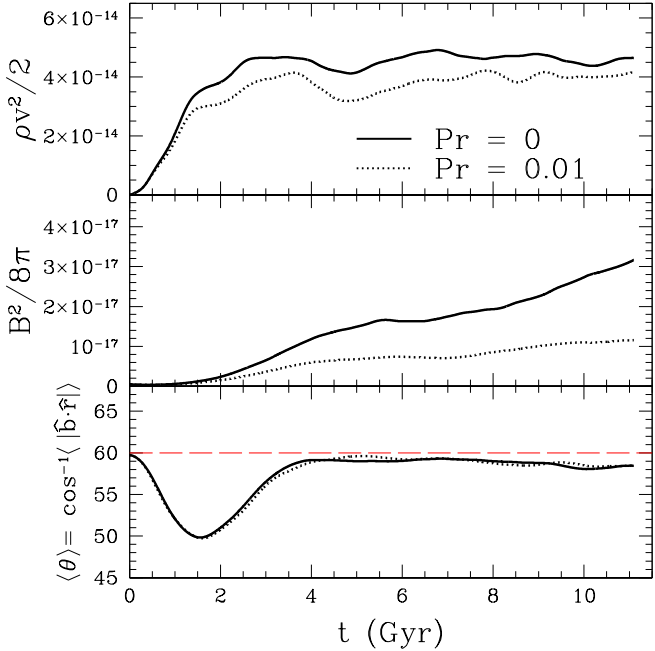}
\caption{The non-linear evolution of volume-averaged properties of the MTI in clusters in our global model with (\textit{dotted lines}) and without (\textit{solid line}) viscosity.  The evolution of the volume-averaged kinetic energy (\textit{top panel}) and the the magnetic field geometry (\textit{bottom panel}) are largely unchanged by the viscosity.  The magnetic geometry is measured with respect to the radial direction:  $\theta = 0 \degree$ corresponds to radial, and $\theta = 60\degree$ (\textit{red, long-dashed line}) corresponds to a random field. The amplification of the volume-averaged magnetic energy (\textit{middle panel}) is modestly suppressed by anisotropic viscosity, similar to what was observed in our local simulations.} \label{fig:mticluster-time}
\end{figure}
\begin{figure}
\centering
\includegraphics[clip=true, scale=0.45]{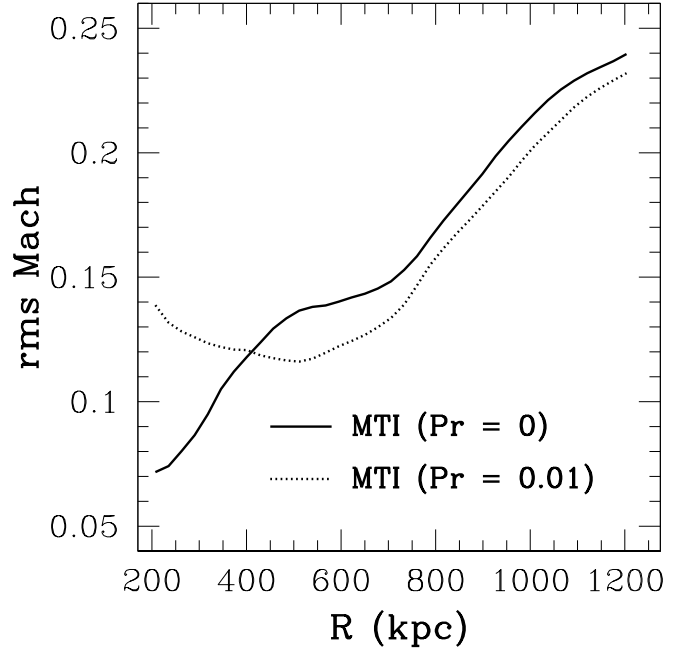}
\caption{The saturated and azimuthally-averaged rms Mach number
  profile due to the MTI in our global model cluster (model MTI in
  Fig. \ref{fig:tctb}).  The inviscid (\textit{solid line}) and
  viscous (\textit{dotted line}) simulations differ only by a small
  amount.} \label{fig:mticluster}
\end{figure}

Figure \ref{fig:mticluster-time} shows the non-linear evolution of the MTI in the global galaxy cluster context.  The evolution of the kinetic energy with time is quite similar for both the viscous and inviscid case just as in the local simulations (Figure \ref{fig:mtilocal}); however, the amplification of the magnetic field is again somewhat suppressed in the viscous case.  The magnetic field stretching necessary to amplify the field is exactly the motion that is damped by anisotropic viscosity.  This damping of magnetic field amplification makes it more difficult to use a turbulent dynamo, regardless of the source of the turbulence, to amplify the magnetic field from primordial values to those observed today in galaxy clusters.  The magnetic geometry evolution is very similar with and without viscosity, as we initially began with a geometrically isotropic magnetic field.  A very slight radial bias exists in the steady state in both cases.  

Figure \ref{fig:mticluster} shows the saturated and azimuthally-averaged rms Mach number profile for our model cluster at a time of 8.3 Gyr.  The turbulence profiles with and without viscosity are almost identical beyond 400 kpc.  More quantitatively, at $r_{500}$ the rms Mach numbers differ by only 5\%.   Near the very center of the cluster, the rms velocity is actually larger for the viscous simulation.  Although we have no definitive explanation, this is perhaps due to the turbulent eddies becoming less aligned with the magnetic geometry.   Since our cluster is on the massive and hot end of the cluster mass function, this cluster has a particularly large value of $\omc/\omb$, which yields the greatest effect of viscosity on the linear MTI (see Figure \ref{fig:dr}).  Since viscosity has little effect on the MTI-generated turbulence in this hot cluster, we expect viscosity to make almost no difference for the MTI in lower mass (and cooler) clusters.
\section{Global Models of the HBI in Cluster Cores}\label{hbicluster}
We now move inwards to consider the effects of viscosity in global models of the cores
of galaxy clusters.  Within 200 kpc, the cooling times are as short as
100 Myr at $\sim 10$ kpc.  These short cooling times, along with
evidence that large mass fluxes of gas are not cooling from a hot to
cold phase constitute the modern cooling flow problem \citep{pf06}.
Namely, what is keeping cool core cluster cools from cooling further?
There has been speculation that thermal conduction from large radii
could offset the cooling luminosity \citep[e.g.,][]{nm01}; however,
the HBI presents a serious problem to this scenario.  Alternatively,
there are many observations of AGN feedback that present a more
plausible heating mechanism.
\begin{figure}
\centering
\includegraphics[clip=true, scale=0.45]{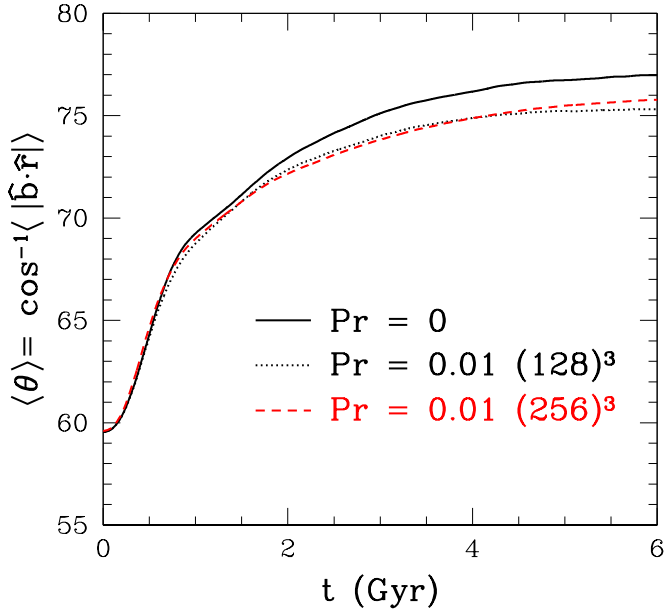}
\caption{Time evolution of the volume-averaged angle of the magnetic
  field with respect to the radial direction in our global cool-core
  cluster model (model CC in Fig. \ref{fig:tctb}).  $\theta=0\degree$
  is radial.  $\theta = 60\degree$ corresponds to a random, isotropic
  magnetic field.  The HBI reorients the magnetic field, reducing the
  effective conductivity, which precipitates the cooling catastrophe
  at 2.4 Gyr.  Very little difference is seen between the inviscid
  (\textit{solid line}) and viscous (\textit{dotted line})
  evolution.  We demonstrate that these results are well-converged by plotting the nearly identical evolution of a viscous simulation with double the resolution (\textit{red dashed line}).} \label{fig:hbicluster-noturb}
\end{figure}

The first cluster model we consider is modeled on observations of Abell 2199
\citep{johnstone02}.  We initialize a cluster in a static NFW halo
with a mass of $3.8\times 10^{14}\;\Mo$ and a scale radius of 390 kpc.
In this potential we calculate a spherically-symmetric atmosphere in
hydrostatic equilibrium and thermal equilibrium with conduction (at
1/3 of Spitzer) exactly balancing cooling.  The model cluster has a
central temperature and electron density of $\simeq 2.0$ keV and
$\simeq 0.021\;\textrm{cm}^{-3}$, respectively, and a temperature and
density of 5 keV and $1.67\times 10^{-3}\;\textrm{cm}^{-3}$ at 200
kpc.  This corresponds to a central cooling time of 1.7 Gyr.  In this
region we initialize a tangled magnetic field with a Kolmogorov power
spectrum and an amplitude of $\la |B| \ra \sim 10^{-8}$.  The magnetic
field is tangled on scales from 50 kpc down to 30 kpc.  The
simulations are computed on a Cartesian grid in a domain extending
from the cluster center to 240 kpc with $(128)^3$ gridpoints.  A
resolution study at $(256)^3$ showed that the behavior of the magnetic
field geometry was almost indistinguishable from the lower resolution
results we focus on here (Fig. \ref{fig:hbicluster-noturb}).  The magnetic field geometry and temperature profiles as a function of radius (not plotted) are nearly identical; as a result, we consider these simulations
very well-converged.  In these simulations, the temperature is fixed to the initial
temperature at a radius of 200 kpc everywhere, but there is no central
boundary condition.  For more details of the set-up see
\citet{pqs09}.  Our CC atmosphere model has $\omc/\omb \sim 10$, which is reasonably consistent with the cool-core ACCEPT cluster plotted in Figure \ref{fig:tctb}.  Relative to observed clusters, our model is somewhat high in $\omc/\omb$ at small radii and on the low-end at larger radii.  This discrepancy is due to the fact that by enforcing thermal equilibrium, we demand that conduction match cooling everywhere without any AGN feedback.  This is, perhaps, a hint that other heating mechanisms, such as AGN feedback, are necessary for real CC clusters.  

We first evolve our fiducial cluster model with anisotropic conduction and cooling only.  We diagnose the effect of the HBI by calculating the mean magnetic field angle from radial $\la \theta\ra = \cos^{-1}\la | \bhat\cdot \hat{\boldsymbol{r}} | \ra$. This quantity is relevant to the thermal evolution as the effective radial thermal conductivity, often called the Spitzer fraction, is given by
\begin{equation}
f_{\textrm{Sp}} \equiv Q_r/\tilde{Q}_r \approx \cos^2 \la | \bhat\cdot \hat{\boldsymbol{r}} | \ra,
\label{eqn:f_sp}
\end{equation}
where $\tilde{Q}_r \equiv - \kappa \dif T/\dif r$, which is the radial
heat flux if the conduction were purely isotropic at the Spitzer
value.  Figure
\ref{fig:hbicluster-noturb} shows that the HBI acts to reorient the
magnetic field to be more azimuthal, thus reducing the effective
radial conductivity.  In this calculation the cooling catastrophe
(defined as the central temperature reaching our imposed floor of 0.05
keV) occurs at $\sim 2.4$ Gyr.

When the effects of Braginskii viscosity are added, the evolution changes very little as shown by the dotted line in Figure \ref{fig:hbicluster-noturb}.  As we saw previously in the local calculations, in 3D interchange-type motions are able to proceed unimpeded by any viscosity at all.  As the HBI drives rather small turbulent velocities ($12\;\kms$ at 2 Gyr), the viscous dissipation does next to nothing to impede the cooling catastrophe which occurs at almost the exact same time as in the inviscid simulation.  In the Braginskii (collisional, anisotropic) limit, heating by viscous dissipation of turbulence is thermally unstable as $\nabla\cdot(\mathsf{\Pi}\cdot \boldsymbol{v}) \propto \nu\propto T^{5/2}$, so even more vigorous turbulence cannot stably balance cooling.

Turbulence driven by galaxy wakes, AGN feedback, structure formation, or any other source can counteract the ability of the HBI to reorient magnetic field lines \citep{mpsq11}.  Quantitatively, turbulence on a scale $L$ with velocity $\delta v$ is able to suppress the HBI when
\begin{equation}
t_{\textrm{eddy}} (L) \simeq \frac{L}{\delta v}\lesssim \xi t_{\textrm{HBI}} \simeq \xi 
\left( g \frac{\dif \ln T}{\dif r} \right)^{-1/2},
\label{eqn:richardson}
\end{equation}
where $t_{\textrm{HBI}}$ is the HBI growth time, and $\xi$ is a dimensionless constant that is determined by simulations \citep{sharma09,pqs10}.   Since $t_{\textrm{eddy}} \propto L^{2/3}$, this inequality is most difficult to satisfy at the outer scale, which dominates the turbulent energy.  Possible sources of turbulence are galaxy wakes \citep{kim05} or AGN feedback itself.
% If we presume five $10^{11} \;\Mo$ galaxies within 200 kpc, and a typical %turbulent driving scale of 40 kpc, then the induced turbulent velocity is %$\delta v \sim 0.08 c_s$.  
\begin{figure}
\centering
\includegraphics[clip=true, scale=0.45]{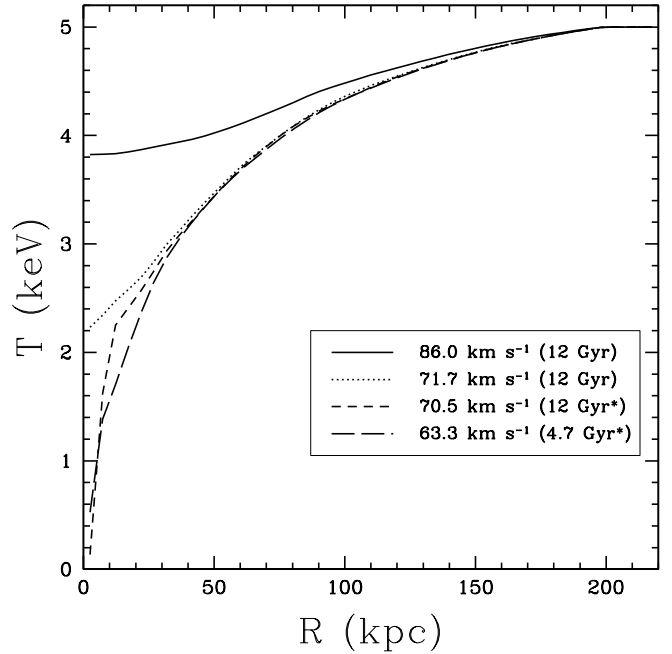}
\caption{Azimuthally-averaged temperature profiles for identical cluster cores with different imposed rms turbulent velocities (see legend) with a fixed driving scale of $L=40$ kpc.  A very strong bimodality is seen in the stability properties of the thermal profiles.  Braginskii viscosity is included.  Driving at $71.7\;\kms$ produces a stable temperature profile; whereas a tiny change in the driving to $70.5\;\kms$ leads to the cooling catastrophe.  The two thermally stable runs report the measured time in parentheses; while the two lower velocity unstable runs (marked with a *) report the time of cooling catastrophe in parentheses.} \label{fig:hbicluster-turb}  
\end{figure}

We simulate the interaction of turbulence, the HBI, and cooling in the
fiducial cluster core model described previously by adding a random
velocity forcing.  We drive the velocity fields in Fourier space with
a flat spectrum on a scale $L= 40\pm 10\;\textrm{kpc}$ such that the
phases are random and the forcing is incompressible.  We have
simulated forcing that results in turbulent velocities from 30--$120\;
\kms$, which corresponds to a pressure support of less than a few
percent in the cluster cores.  The turbulent energy input from this
forcing. $\dot{e}_{\textrm{turb}} \simeq \rho (\delta v)^3/L$, is
negligible compared to the cooling rate, and thus the turbulence is
not energetically important. 

Figure \ref{fig:hbicluster-turb} shows the late-time, azimuthally-averaged temperature profiles for the same cluster model with different injected turbulent velocities.  Viscosity is present at the Spitzer value for these simulations.  At low turbulent velocities, $\delta v \lesssim 70.5\;\kms$, the HBI acts to reorient the magnetic field lines, reducing the effective thermal conductivity and driving the cluster core towards a cooling catastrophe.  The time of the cooling catastrophe is reported in parentheses in the legend.  Turbulence does indeed delay the cooling catastrophe somewhat, but below a critical value is incapable of preventing it.  For example, a turbulent forcing velocity of $70.5\;\kms$ is able to delay the cooling catastrophe from 2.4 Gyr with no turbulence to 12 Gyr with turbulence. 

On the other hand, for larger turbulent velocities, the turbulence is
able to keep the magnetic field geometry close to isotropic. In this
case, the central temperature approaches a new thermally stable state
at a higher temperature.  In fact, as $\delta v$ increases, the
central temperature increases due to both conduction and the larger
radially-inward advective heat flux driven by the turbulence.  In the final state of
the cluster, the central entropy is significantly larger than in the
initial condition.  For example, a driving velocity of $\delta v
\simeq 86.0\;\kms$ takes the initial central entropy from $K_0 \simeq
26.5 \;\kevcm$ to $90\;\kevcm$.

We find a strikingly strong bimodality in the behavior of the cluster temperature profiles and central entropies.   Figure \ref{fig:hbicluster-turb} shows that changing the driving velocity by a mere $1.2\;\kms$ causes a cluster to go from heating approximately balancing cooling ($71.7\;\kms$) to a core experiencing a cooling castatrophe ($70.5\;\kms$).  The former simulation is almost exactly on the border of  the stability boundary and departs only marginally from the initial temperature profile.  This particular case where conduction plus turbulence is balancing cooling to maintain a cool core is an example of fine tuning.  For turbulent velocities 5\% above or below this value, the core evolves to a clear non-cool-core (NCC) profile or experiences a cooling catastrophe.  Note, that we have not included any source of AGN feedback that could avert the cooling catastrophe in the latter situation.

What does Braginskii viscosity do?  Including the viscosity makes
little difference to the fundamental bimodality observed.  In fact, it
may even sharpen the transition since the viscosity scales as
$T^{5/2}$.  A cluster whose core is heating from the initial condition
becomes more viscous and thus a given turbulent velocity is more
readily able to reorient the magnetic field relative to the HBI as the temperature increases.

We checked the robustness of our conclusions by varying our initial fiducial physical a variety of ways:
\begin{itemize}
\item \textit{A hotter, more massive CC cluster}.  We initialize a model with a central temperature of 2.5 keV and entropy of $K_0 = 27\,\kevcm$ which increases to a temperature of 8 keV at 300 kpc.   In the absence of turbulence this cluster proceeds to a cooling catastrophe in $\sim 2.1$ Gyr, much as our fiducial model.  With turbulent driving at an rms velocity of $57\,\kms$, the cluster transitions to a NCC cluster with a central entropy of $K_0 = 71\,\kevcm$.  The viscous and inviscid cases are very similar as in our fiducial calculation.  
\item \textit{A NCC cluster of equivalent mass}.  We initialize our
  fiducial model with temperature ranging from 4 keV to 5 keV over 200
  kpc and a central entropy of $98\,\kevcm$.    From linear theory, one expects viscosity to reduce the HBI
  growth rates somewhat due to the higher ratio of $\omc/\omb$
  relative to our fiducial model (see Figure \ref{fig:dr}); however, the evolution of the temperature profile is almost indistinguishable with and without viscosity.
  Without any imposed turbulence this cluster reaches a cooling
  catastrophe around 6.2 Gyr which is somewhat faster than the
  initially predicted cooling time of 7.2 Gyr.\footnote{This cooling
    time is an overestimate as the Bremsstrahlung cooling rate
    increases $\propto T^{-3/2}$  as material cools at constant pressure.}  
\item \textit{A hotter, more massive NCC cluster}  We initialize a hotter cluster with a mass of $5.2\times 10^{15} \; \Mo$ and a temperature that ranges from 6 keV at the center to 8 keV at 300 kpc.   This model has a scale radius of 650 kpc and a central entropy of $K_0 = 129\;\kevcm$.  The profile of $\omc/\omb$ is shown in Figure \ref{fig:tctb} with the NCC (HBI) label.   The value of $\omc/\omb$ is $\sim 100$ at 50 kpc and $\sim 40$ at 100 kpc, consistent with the ACCEPT NCC measurements.  Figure \ref{fig:hbicluster-ncc} shows the evolution of the radial profile of the magnetic field angle from its initial statistical isotropy of $60\degree$.  In the majority of the volume, the magnetic field is reoriented to be substantially more tangential ($\theta \gtrsim 75\degree$) with little difference between the viscous and inviscid cases.  The small radial bias around 20 kpc is caused by inhomogeneous radial infall as the cluster begins to suffer a cooling catastrophe.  Based on the profiles in Figure \ref{fig:tctb}, viscosity is even more important at these small radii in the initial cluster model; additional suppression of the HBI may be present but is overwhelmed by the effects of cooling in our global calculations.  At larger radii, viscous effects are smaller, and the HBI has already reduced conduction from the outskirts.  Overall this hot, NCC cluster model shows that the HBI can operate even when viscous effects are important.  
\begin{figure}
\centering
\includegraphics[clip=true, scale=0.45]{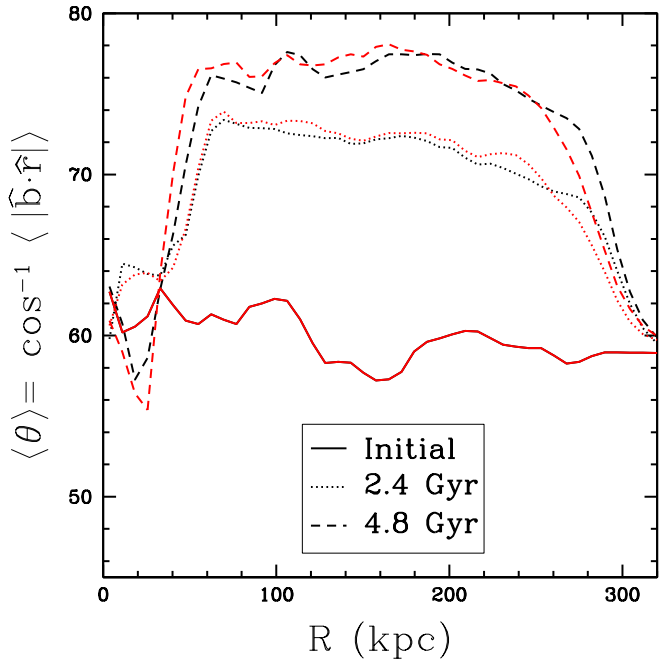}
\caption{Radial profiles of the azimuthally-averaged angle of the magnetic
  field with respect to the radial direction in our global non-cool-core
  cluster model (model NCC in Fig. \ref{fig:tctb}) at two times.  $\theta=0\degree$
  is radial.  $\theta = 60\degree$ corresponds to a random, isotropic
  magnetic field.  The HBI reorients the magnetic field, reducing the
  effective conductivity, which precipitates the cooling catastrophe
  at 2.4 Gyr.  Very little difference is seen between the inviscid
  (\textit{black curves}) and viscous (\textit{red curves})
  evolution.  The radial bias seen around 20 kpc is due to inhomogeneous radial infall.} \label{fig:hbicluster-ncc}  
\end{figure}
\item \textit{Reduced magnetic field correlation length}.  We simulate our fiducial cluster with the magnetic field tangled on scales ranging from 20 kpc to 10 kpc.  The shorter correlation length can reduce the effective conduction length scale, increasing $\omc$, and reducing the HBI growth rate.  The thermal evolution, however, is nearly identical to our fiducial model both with and without viscosity.
\item \textit{Higher initial magnetic field}.  We simulate our fiducial cluster with the magnetic field increased to $\la |B| \ra \sim 10^{-6}$ which is similar to observed magnetic field strengths \citep{ct02}.  Magnetic tension is able to suppress short wavelength modes in a way very similar to viscosity.  With the higher field, the HBI proceeds slightly more slowly and the fiducial cluster reaches a cooling catastrophe approximately 200 Myr later.  We find no difference between the viscous and inviscid simulations.  
\end{itemize}

\section{Discussion and Conclusions}\label{sec:conclusions}

The dilute plasma in the ICM of galaxy clusters has a mean free path
along magnetic field lines that is many orders of magnitude larger
than the electron/proton gyroradii.  As a result, thermal conduction
by electrons and momentum transport by ions are strongly anisotropic
with respect to the magnetic field.  In this paper we have carried out
the first simulations of buoyancy instabilities in the ICM (the MTI
and HBI) that include anisotropic conduction and viscosity
simultaneously and self-consistently.
This extends the analytic work of K11, who showed that viscosity can
change the instabilities in the limit of very rapid
conduction/viscosity, suppressing the growth rates of some of the
modes (Fig. \ref{fig:dr}).
  
In 2D simulations of the HBI, viscosity changes the nonlinear
saturation of the instability.  Figure \ref{fig:hbi2d} shows that strong viscosity drives the magnetic field to bunch
up in prominent vertical structures.  However, in 3D the magnetic
field lines slip past each other in interchange-like motions.  These
interchange motions are not damped by parallel viscosity. As a result,
in 3D the saturated state of the HBI (as measured by statistical
quantities like $\la | \hat{b}_z | \ra$) is not as strongly altered by
the addition of viscosity, although it can take longer to
reach the saturated state because viscosity slows the growth of the
modes (Figure \ref{fig:hbi2d3dcomp}).  This conclusion is true even for very
rapid conduction (and thus very rapid viscous damping), with $\omc(L) \sim
50 \, \omb$ (where the conduction frequency here is defined across the
scale of the box $L$); this is a value characteristic of cluster cores
(Fig. \ref{fig:tctb}).

In the cores of galaxy clusters the cooling times are short compared
to the age of the universe.  There is a long-standing problem of
finding processes that can provide sufficient heating to counteract
this cooling.  One possible solution is for conduction to bring in
heat from the large thermal reservoir beyond the core.  However, in
the absence of other sources of turbulence, the HBI exacerbates the
cooling flow problem by reorienting magnetic field lines to be more
azimuthal, thus reducing the effective radial thermal conductivity,
$f_{\textrm{Sp}}$.  Our global simulations of realistic cool-core
cluster models (Fig. \ref{fig:tctb}) demonstrate that this
reorientation of the magnetic geometry by the HBI is only minimally
influenced by anisotropic viscosity (Fig.
\ref{fig:hbicluster-noturb} \& \ref{fig:hbicluster-ncc}).

We also simulated galaxy cluster cores with additional turbulent
forcing, intended to mimic the turbulence driven by galaxy wakes
and/or AGN feedback.  Turbulence is capable of suppressing the
magnetic field reorientation of the HBI when the eddy turnover time is
similar to the instability growth time \citep[see Figure 11 of
][]{mpsq11}.  The interplay between turbulence, the HBI, and cooling
in the cluster core, results in a very strong bimodality in the
temperature profile and other cluster properties such as the central
entropy \citep{pqs10}.  Modest levels of subsonic turbulence, less
than $100\;\kms$, are capable of transforming a cool core cluster into
a non-cool-core cluster.  This turbulence is energetically small
compared to the cooling luminosity in the cluster core; instead, the
turbulence can be considered a catalyst that enables conduction to
efficiently couple to the central regions.  Larger turbulent
velocities result in larger central temperatures and entropies, aided
by both conduction and an inwardly-directed turbulent heat transport.
With Braginskii viscosity, these conclusions about cluster bimodality
are unchanged.  In fact, we believe that viscosity aids the generation
of a bimodal cluster population: as the cluster core heats up, it
becomes more viscous, which makes it somewhat more difficult for the
HBI to re-orient the magnetic field, further promoting the transition
to a non-cool-core cluster.

Our numerical experiments are encouragingly consistent with the
observed bimodality in cluster core properties \citep{voit08, cav09}.
We find that cool-core clusters (with central entropies $K_0 \lesssim
30\;\kevcm$) cannot be efficiently heated by conduction and must have
an additional source of heating.  This source is plausibly AGN feedback
through bubbles, jets, or other coupling mechanisms.  Indeed,
cool-core clusters generically show evidence of radio emission
consistent with AGN activity.
% H$\alpha$ emission, and brightest cluster galaxy color
%gradients that are all suggestive of ICM material cooling and
%producing AGN feedback. 
With turbulence above a critical threshold value, however, we find
that conduction is able to transform a cool-core cluster into a
non-cool-core cluster with a much higher central entropy---this does
not necessarily require AGN activity though it does require roughly
volume-filling turbulence.
% The turbulent energy itself is small
%compared to either the cooling luminosity or conductive energy input.
Observationally, this is consistent with the lack of significant AGN feedback
indicators for high central entropy clusters.

In local Cartesian simulations of the MTI we find that the evolution of kinetic energy is statistically unchanged by anisotropic viscosity; however, the amplification of the magnetic field is diminished with the addition of viscosity (Fig. \ref{fig:mtilocal}).  Magnetic fields are amplified by turbulence increasing the length of magnetic field lines; this motion is precisely the one damped by anisotropic viscosity.  As a result, the magnetic field is amplified more slowly than the kinetic energy with anisotropic viscosity.  For simulations with initially horizontal fields, increased viscosity leads to a small vertical bias in the magnetic field direction relative to isotropy.  Simulations that start with initially vertically magnetic fields, a nonlinearly unstable configuration, evolve much more slowly at high viscosity (bottom panel of Fig. \ref{fig:mtilocal}).  This mechanism explains the small vertical bias seen for higher viscosity.   

In the outskirts of galaxy clusters, the temperature declines with
increasing radius and the plasma is unstable to the MTI rather than
the HBI.  At these radii, the MTI can generate significant turbulent
velocities up to an rms Mach number of $\la \mathcal{M}\ra\sim 0.35$
near $r_{200}$.  This turbulence represents a non-thermal source of
pressure that can bias hydrostatic estimates of cluster masses.  As
emphasized in Parrish et al. (2011), this turbulence is in addition to
that generated by structure formation.  In the global cluster models of the MTI presented here (Fig.
\ref{fig:mticluster}), we find that viscosity only weakly affects this
turbulence, as predicted analytically by K11.  The rms velocities with
viscosity are only 5\% smaller than the inviscid case.  Consistent with our local simulations, we find that the amplification of the magnetic field is modestly reduced for simulations with viscosity.  The fact that anisotropic viscosity specifically damps all sources of turbulence that couple to magnetic field amplification makes it more difficult to invoke turbulent mechanisms for amplifying the magnetic field in clusters from primordial values. 

Although our results with anisotropic viscosity do not significantly change any of
the previous conclusions about the role of the HBI and MTI in the ICM,
we believe that it is important to have established this conclusively.
Most importantly, a fluid with a low collisionality like the ICM or
the solar wind has a high viscosity, as the viscosity scales inversely
with the collision frequency; the Reynolds number is correspondingly
rather small (Fig. \ref{eqn:Reynolds}). Thus the importance of
viscosity is not a priori clear and needed to be explicitly studied.
In future work it will be interesting to consider the role of
anisotropic viscosity on other problems in the ICM, such as cold
fronts or the large surface brightness fluctuations (ripples) observed in Perseus \citep{fab06}.
\appendix
\section{Algorithm and Tests of Anisotropic Viscosity}
\subsection{Method} \label{appendix:method}
We now describe the details of  our algorithm for anisotropic momentum transport.  This algorithm is similar to that used in \citet{dongstone09}.  For simplicity consider the momentum and energy equations with only anisotropic viscosity 
\begin{eqnarray}
\frac{\partial (\rho \boldsymbol{v})}{\partial t} + \nabla \cdot \mathsf{\Pi} & = & 0,\\ \label{eqn:simple_mass}
\frac{\partial E}{\partial t} + \nabla \cdot\left(\mathsf{\Pi}\cdot\boldsymbol{v}\right) & = & 0,\label{eqn:simple_energy}
\end{eqnarray}
where the viscous stress tensor is given by Equation
\ref{eqn:viscosity_tensor}.  We proceed to discretize this in a very
similar manner to how anisotropic conduction was first discretized in
\citet{ps05}. For example, the first component in square brackets of
the tensor has terms that can be written as
\begin{equation}
\bhat\bhat:\nabla \boldsymbol{v} = \bhat_x\left(\bhat_x\frac{\partial v_x}{\partial x} + \bhat_y\frac{\partial v_x}{\partial y}
+ \bhat_z\frac{\partial v_x}{\partial z} \right) +\bhat_y\left(...\right) +\bhat_z\left(...\right),
\label{eqn:stress-example}
\end{equation}
where the ellipsis indicates similar derivatives of $v_y$ and $v_z$, respectively.   We will see momentarily that this stress is computed as an area-averaged, face-centered quantity which then requires velocity derivatives to be appropriately centered.  

The calculation of $\bhat_x^2 \partial v_x/\partial x$ is straightforwardly discretized from the cell-centered velocities.  The calculation of the transverse velocity gradients, such as $\bhat_x \bhat_y \partial v_x/\partial y$ requires special care as monotonicity is required.  This issue surfaces in anisotropic thermal conduction when it proves essential to limit the slope of the transverse temperature derivatives, e.g. the term $\bhat_x \bhat_y \partial T/\partial y$ in the calculation of $q_x$, to prevent heat from being conducted from cold to hot regions \citep{sh07}.  
\begin{figure}
\centering
\includegraphics[clip=true, scale=1.0]{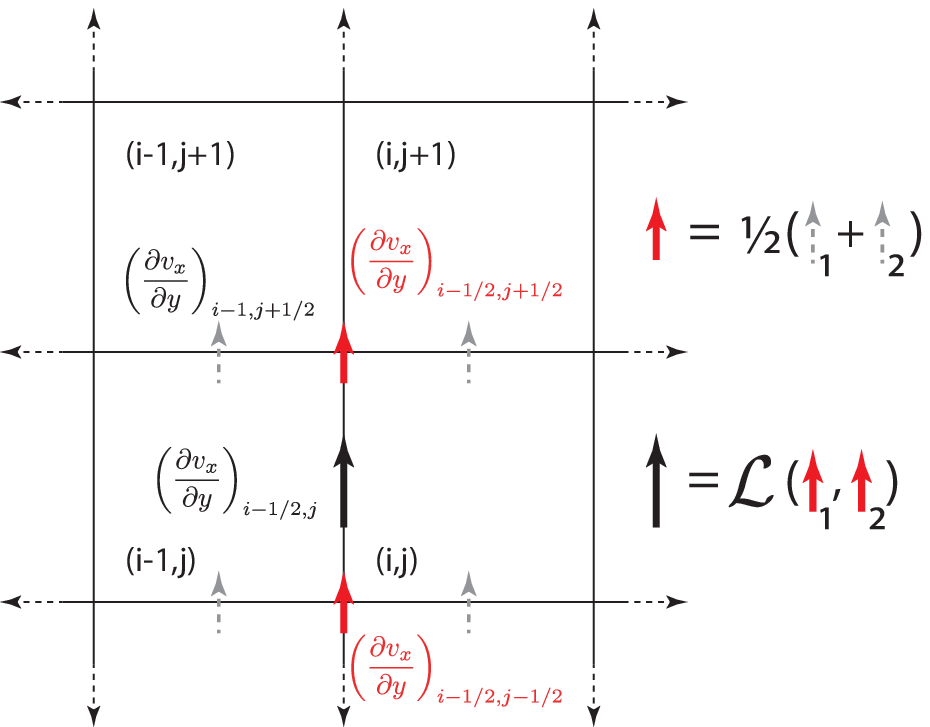}
\caption{This figure illustrates the centering of quantities necessary for calculating the the shear on the $x$-face of cell ($i,j$) due to the transverse velocity gradient, $\partial v_x/\partial y$.  The velocity derivatives at cell corners (\textit{red arrows}) are simple averages in Equation \ref{eqn:velocity-corner}.  These derivatives are then averaged with a limiter ($\mathcal{L}$) using Equation \ref{eqn:velocity-limiter} to get the face-centered shear (\textit{thick black arrows}) used for calculating the stress. \label{fig:grid}  }
\end{figure}
This type of non-monotonicity can lead to negative internal energies with conduction.  The same symptom can occur with viscosity, and it is necessary to ensure that momentum flows in the correct direction at all times.  We therefore interpolate the transverse velocity gradients to cell faces with a slope limiter:
\begin{equation}
\left(\frac{\partial v_x}{ \partial y}\right)_{i-\onehalf, j} = \mathcal{L} \left[ \left(\frac{\partial v_x}{\partial y}\right)_{i-\frac{1}{2}, j + \onehalf},  
\left(\frac{\partial v_x}{\partial y}\right)_{i-\onehalf, j - \onehalf}\right]
\label{eqn:velocity-limiter}
\end{equation}
where $i-\onehalf$ represents the left face of cell ($i,j$), and $\mathcal{L}$ represents a slope limiter.  The velocity gradients at cell edges are simply arithmetic averages, e.g.
\begin{equation}
\left(\frac{\partial v_x}{\partial y}\right)_{i-\frac{1}{2}, j + \onehalf} = \onehalf
\left[ \frac{v_{x, i-1, j+1} - v_{x,i-1, j}}{\Delta y}  +  \frac{v_{x, i, j+1} - v_{x,i, j}}{\Delta y}  \right].
\label{eqn:velocity-corner}
\end{equation}
We choose the monotonized central difference (MC) limiter for our slope limiter \citep[see][for details]{vanLeer}. 

Our explicit methods for anisotropic conduction and viscosity both have timesteps that are more restrictive than the MHD timestep as they are parabolic equations.  The momentum diffusion has  a Courant-limited timestep that is proportional to $(\Delta x)^2$.  Therefore, we choose to sub-cycle both diffusive operators with respect to the MHD timestep.  For the anisotropic viscosity, we are able to simply call the standard MHD boundary conditions in between each sub-cycle.   Due to the smaller physical diffusivity, we are able to take much longer timesteps for viscosity than conduction.  Sub-cycling the conduction and viscosity do not change the evolution of the HBI compared to reducing the MHD timestep to match the conduction timestep. 

After these preliminaries, we are now prepared to outline the method for anisotropic momentum transport.  Careful attention is paid to the centering of the terms in the difference equations.  Our method resembles the underlying Godunov scheme for Athena in that conservative cell-centered variables, e.g. momentum, are updated by differencing face-centered fluxes, e.g. the viscous stress---such a scheme is manifestly conservative.  Our method for calculating one sub-cycle of a cell centered on ($i,j$) is as follows:
\begin{enumerate}
\item Apply the MHD boundary conditions to synchronize all ghost zones across processors.
\item Calculate face-centered velocity gradients, including limiters on transverse gradients, e.g. $(\partial v_x/ \partial y)_{i-\onehalf, j}$.
\item Calculate the face-centered viscous prefactor
\begin{equation}
\left(\rho \nu\right)_{i-\onehalf, j} = \onehalf \left\{ \left[ \rho \nu(T,\rho) \right]_{i-1, j} +  \left[ \rho \nu(T,\rho) \right]_{i+1, j}\right\} ,
\end{equation}
where the diffusion coefficient (sometimes known as the kinematic viscosity) depends on density and temperature as $\nu\propto T^{5/2}\rho^{-1}$.  The density dependence drops out of the stress tensor.  The density dependence, however, is relevant for calculation of the viscous timestep. 
\item Calculate the face-centered viscous stresses, e.g. $\mathsf{\Pi}_{i-\onehalf, j}$ on the left $x$-face.
\item Difference the viscous stresses to get a cell-centered update to the energy and momentum equation, which is in the $x$-direction:
\begin{eqnarray}
\Delta\left(\rho v_x\right)_{i,j} &=& \frac{\Delta t}{\Delta x}
\left(\mathsf{\Pi}_{i+1/2,j} - \mathsf{\Pi}_{i-1/2,j}\right),\\
\Delta\left(E\right)_{i,j} &=& \frac{\Delta t}{\Delta x}
\left[ \left(\mathsf{\Pi}\cdot \boldsymbol{v}\right)_{i+1/2,j} - 
  \left(\mathsf{\Pi}\cdot\boldsymbol{v}\right)_{i-1/2,j}\right].
\end{eqnarray}
There are similar terms in the $y$- and $z$-directions.  The $\Delta t$ here is the sub-cycle timestep.
\end{enumerate}
Steps (i)--(v) are repeated for each sub-cycle.
\subsection{Verification}\label{appendix:verification}
It is important to verify a new numerical method, such as anisotropic viscosity, with test problems with known solutions.  To wit, we begin with two tests of the physics of linear MHD waves.  We take advantage of the property that Braginskii viscosity damps viscous motions only along field lines; however, our numerical implementation will have some spurious $\nu_{\perp}$.  

We test our algorithm by measuring the viscous damping rates of MHD waves.
These damping rates are sensitive to both the magnitude and anisotropy of the
viscosity and therefore provide a stringent test of our code.  Moreover, when
the viscosity is small enough (quantified later), the damping rates can be
calculated analytically, permitting a rigorous point of comparison for our
simulations.  In order to test the multi-dimensional nature of the code, we
pick the wave vector $\boldsymbol{k}$ to be 45$\degree$ from the grid axes and the
magnetic field $\boldsymbol{B}$ to be 56.3$\degree$ from $\boldsymbol{k}$.  Thus, nothing is
aligned with the grid axes and all terms in our algorithm are evaluated.

Alfv\'en waves directly probe the anisotropy of our algorithm.  The
fluid motions induced by these waves are purely transverse to the
magnetic field and therefore are not damped by parallel viscosity.
Any measured damping represents a spurious perpendicular (or
isotropic) viscosity due to errors in the algorithm.  We confirm
that our code does not damp Alfv\'en waves: even with an
\emph{extremely} large viscosity $\nu = 1 \lambda^2 / T$, where
$\lambda$ is the wavelength and $T$ is the wave period, we measure a
damping rate of $1.8 \times 10^{-4} / T$.  This damping is only $\sim
40$\% larger than that produced by the inviscid, default version of
Athena with no explicit viscosity.  For more reasonable viscosities, this damping is
much reduced.  Thus, the spurious, cross-field diffusion in our
algorithm is typically much smaller than other sources of numerical
dissipation in our simulations.

% Convergence Plot
\begin{figure*}
  \includegraphics[width=5in]{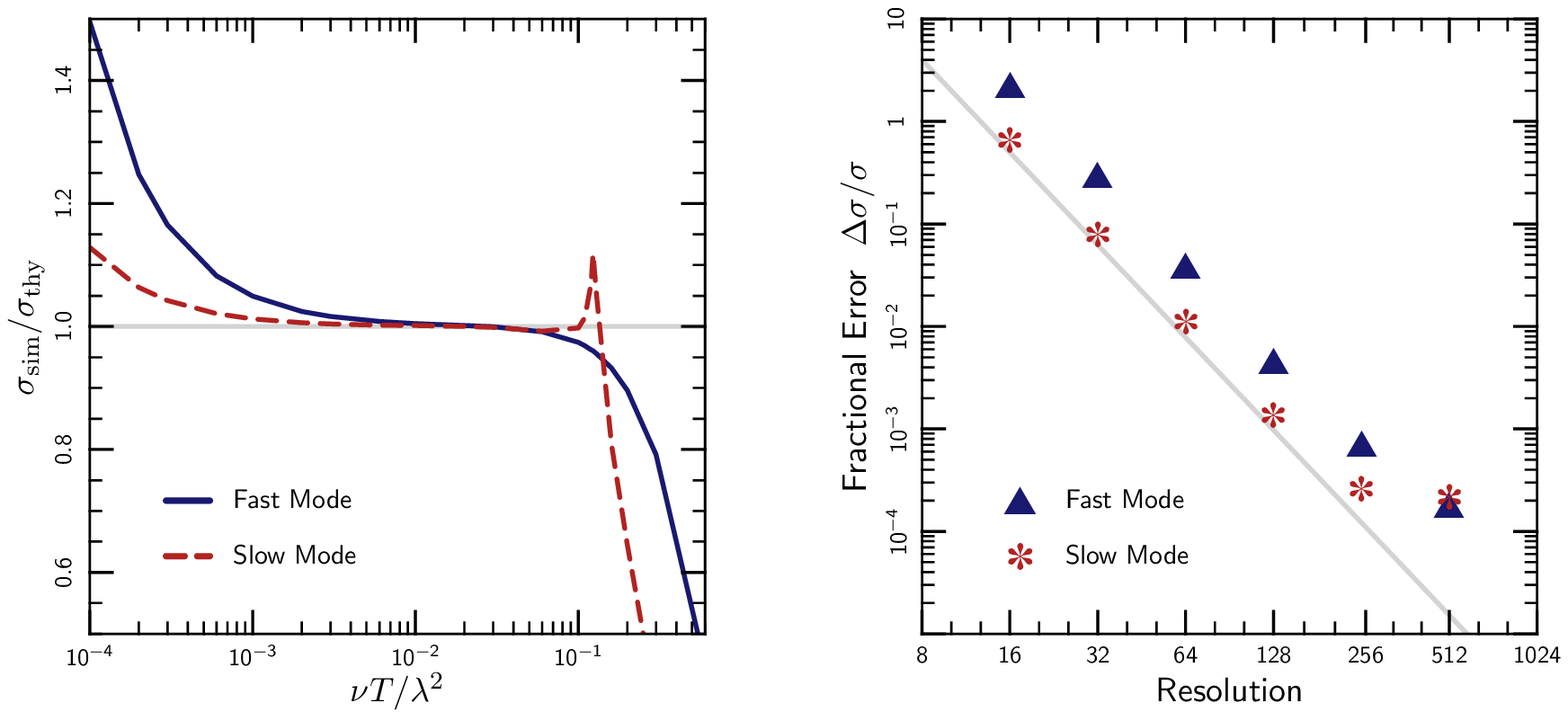}
  \caption{Measurement of the damping rates $\sigma$ of the fast and
    slow magnetosonic modes.  \textit{(Left:)} The ratio of measured
    to predicted damping rates as a function of viscosity (in units of
    wavelength\textsuperscript{2} / wave period).  The agreement is
    very good over $\sim 2$~orders of magnitude in $\nu$.  The
    disagreement at low $\nu$ is likely caused by numerical viscosity
    in other parts of Athena's algorithm, while the disagreement at
    large $\nu$ represents a breakdown in our linear damping
    calculation of $\sigma$. \textit{(Right:)} The fractional error in
    the analytic damping rates for our fiducial viscosity, $\nu
    T/\lambda^2 = 0.01$, as a function of resolution.  The grey line
    represents third-order convergence.  This excellent convergence
    result is the convolution of the second order accuracy of the
    linear waves themselves with the second order accuracy of our
    viscous transport algorithm. }%
  \label{fig:convergence}
\end{figure*}
We test the parallel component of the viscosity by measuring the damping of
fast and slow magnetosonic waves.  Both of these waves have a component of velocity
parallel to the magnetic field and thus damp at a rate
\begin{equation}
\sigma = - \frac{\nu k^2}{6} \left[ 
  (\khat\cdot\vhat) 
  - 3 (\bhat\cdot\khat) (\bhat\cdot\vhat) \right]^2,
\label{eqn:ms-damp}
\end{equation}
where $\bhat$, $\khat$, and $\vhat$ are unit vectors in the directions
of the magnetic field, wave vector, and wave perturbation velocity,
respectively \citep{brag65}.  Figure~\ref{fig:convergence} compares
the damping rates measured in our simulations with this analytic
expectation.  In the left panel, we plot the ratio of the measured to
expected damping rates.  The disagreement at low viscosities is most
likely caused by the numerical, isotropic viscosity inherent in
Athena's integrator.  This viscosity is not included in the analytic
expectation, so equation~\ref{eqn:ms-damp} underestimates the damping
rate in this limit.  With higher viscosity $\nu \gtrsim 0.1$, the
linear damping assumption in the derivation of equation~\ref{eqn:ms-damp} is not valid.  The algorithm
performs well in between these limits.

In order to test the the numerical convergence of our algorithm, we
chose a fiducial viscosity $\nu = 0.01 \lambda^2 / T$ , such that the
damping remains small but is not dominated by numerical viscosity.  We
plot the fractional error in the damping rate $\Delta \sigma / \sigma
\equiv (\sigma_{\mathrm{meas}} - \sigma_{\mathrm{theory}}) /
\sigma_{\mathrm{theory}}$ in the right panel of
Figure~\ref{fig:convergence}.  This is not the usual L2 norm that is
often plotted for linear wave convergence tests.  Even at 16
zones/wavelength, the damping rate is correct to order unity.  The
quantity $\Delta \sigma/\sigma$ converges at third order.  This
excellent convergence results from a convolution of the increasing
accuracy of the linear waves themselves as well as the improved
accuracy of the viscous transport, both of which are second-order
algorithms.

Perhaps the most demanding test of our combined algorithm for MHD,
anisotropic conduction, and anisotropic viscosity is our measurement
of the linear growth rates of the HBI and MTI in Figure~\ref{fig:dr}.
We are able to confirm the analytic growth rates to within $\sim 1$\%,
but unfortunately cannot reach the same accuracy as with the linear
MHD waves.

At least two effects limit our precision when measuring the growth
rates of the HBI and MTI.  First, the analytic growth rates to which
we compare are only strictly known in the WKB limit $k H \gg 1$.
Practical considerations limit our simulation setups to $k H \sim
100$, however.  We expect corrections to the analytic growth rates of
order $1/(k H) \sim 1\%$, similar the typical discrepancy in our
measurements.  Additionally, the structure of the eigenmodes of the
HBI and MTI depend on the growth rate.  Since we only know the growth
rate to an accuracy of $\sim 1$\%, we cannot initialize the simulation
in an exact eigenstate.  Thus, the perturbations we apply do not grow
strictly exponentially in our simulations; there is an initial period
of order $t_{\mathrm{buoy}}$ during which the perturbations settle
into their respective eigenstates. 
In practice, this limits the period of growth from $\sim 2
t_{\mathrm{buoy}}$, when exponential growth begins, to $\sim 4
t_{\mathrm{buoy}}$, when the instability begins to become nonlinear.  Thus we
have only a limited time window over which to fit the data and cannot reach
an arbitrary level of precision.  It might seem that we could delay
saturation, and therefore improve the accuracy of our growth rate
measurement, by using a smaller initial perturbation.  In practice,
however, our boundary conditions do not hold hydrostatic equilibrium
perfectly and eventually generate motions of order $5\times10^{-5}
c_{\mathrm{s}}$.  Thus, when we use a weaker perturbation we lose it
in the noise.

Despite these limitations, we are able to confirm the growth rates
with a comparable accuracy to the validity of the analytic theory.
This, along with our measurements of the damping of linear MHD waves,
represents a strong validation of our code.

\acknowledgments We thank Jim Stone for sharing the anisotropic viscosity module used in Dong \& Stone (2009) with us and for useful feedback on the paper.  We also thank Matt Kunz for helpful comments on an early draft.  Support was provided in part by NASA Grant ATP09-0125, NSF-DOE
Grant PHY-0812811, and by the David and Lucille Parker Foundation.
Support for P.S. was provided by NASA through the Chandra Postdoctoral
Fellowship grant PF8-90054 awarded by the Chandra X-Ray Center, which
is operated by the Smithsonian Astrophysical Observatory for NASA
under contract NAS8-03060.  We would like to thank the hospitality of
the KITP where much of this work was performed and supported in part
by the National Science Foundation under grant PHY05-51164.  The
computations for this paper were perfomed on the \textit{Henyey}
cluster at UC Berkeley, supported by NSF grant AST-0905801 and through
computational time provided by the National Science Foundation through
the Teragrid resources located at the National Institute for
Computational Sciences under grant TG-AST080049.

\bibliography{cluster}

\begin{thebibliography}{32}
\expandafter\ifx\csname natexlab\endcsname\relax\def\natexlab#1{#1}\fi

\bibitem[{{Balbus}(2000)}]{bal00}
{Balbus}, S.~A. 2000, \apj, 534, 420

\bibitem[{{Bogdanovi{\'c}} {et~al.}(2009){Bogdanovi{\'c}}, {Reynolds},
  {Balbus}, \& {Parrish}}]{bog09}
{Bogdanovi{\'c}}, T., {Reynolds}, C.~S., {Balbus}, S.~A., \& {Parrish}, I.~J.
  2009, \apj, 704, 211

\bibitem[{Braginskii(1965)}]{brag65}
Braginskii, S.~I. 1965, Reviews of Plasma Physics, ed. M.~A. Leontovich, Vol.~1
  (New York: Consultants Bureau), 205

\bibitem[{{Carilli} \& {Taylor}(2002)}]{ct02}
{Carilli}, C.~L., \& {Taylor}, G.~B. 2002, \araa, 40, 319

\bibitem[{{Cavagnolo} {et~al.}(2009){Cavagnolo}, {Donahue}, {Voit}, \&
  {Sun}}]{cav09}
{Cavagnolo}, K.~W., {Donahue}, M., {Voit}, G.~M., \& {Sun}, M. 2009, \apjs,
  182, 12

\bibitem[{{Dong} \& {Stone}(2009)}]{dongstone09}
{Dong}, R., \& {Stone}, J.~M. 2009, \apj, 704, 1309

\bibitem[{{Fabian} {et~al.}(2006){Fabian}, {Sanders}, {Taylor}, {Allen},
  {Crawford}, {Johnstone}, \& {Iwasawa}}]{fab06}
{Fabian}, A.~C., {Sanders}, J.~S., {Taylor}, G.~B., {Allen}, S.~W., {Crawford},
  C.~S., {Johnstone}, R.~M., \& {Iwasawa}, K. 2006, \mnras, 366, 417

\bibitem[{{Gardiner} \& {Stone}(2008)}]{gs08}
{Gardiner}, T.~A., \& {Stone}, J.~M. 2008, Journal of Computational Physics,
  227, 4123

\bibitem[{{Johnstone} {et~al.}(2002){Johnstone}, {Allen}, {Fabian}, \&
  {Sanders}}]{johnstone02}
{Johnstone}, R.~M., {Allen}, S.~W., {Fabian}, A.~C., \& {Sanders}, J.~S. 2002,
  \mnras, 336, 299

\bibitem[{{Kim} {et~al.}(2005){Kim}, {El-Zant}, \& {Kamionkowski}}]{kim05}
{Kim}, W., {El-Zant}, A.~A., \& {Kamionkowski}, M. 2005, \apj, 632, 157

\bibitem[{{Kunz}(2011)}]{kunz11}
{Kunz}, M.~W. 2011, \mnras, 1288

\bibitem[{{McCourt} {et~al.}(2011{\natexlab{a}}){McCourt}, {Parrish}, {Sharma},
  \& {Quataert}}]{mpsq11}
{McCourt}, M., {Parrish}, I.~J., {Sharma}, P., \& {Quataert}, E.
  2011{\natexlab{a}}, \mnras, 413, 1295

\bibitem[{{McCourt} {et~al.}(2011{\natexlab{b}}){McCourt}, {Sharma},
  {Quataert}, \& {Parrish}}]{msqp11}
{McCourt}, M., {Sharma}, P., {Quataert}, E., \& {Parrish}, I.~J.
  2011{\natexlab{b}}, ArXiv:1105.2563

\bibitem[{{Narayan} \& {Medvedev}(2001)}]{nm01}
{Narayan}, R., \& {Medvedev}, M.~V. 2001, \apjl, 562, L129

\bibitem[{{Parrish} {et~al.}(2011){Parrish}, {McCourt}, {Quataert}, \&
  {Sharma}}]{pmqs11}
{Parrish}, I.~J., {McCourt}, M., {Quataert}, E., \& {Sharma}, P. 2011, \mnras,
  L356

\bibitem[{{Parrish} \& {Quataert}(2008)}]{pq08}
{Parrish}, I.~J., \& {Quataert}, E. 2008, \apjl, 677, L9

\bibitem[{{Parrish} {et~al.}(2009){Parrish}, {Quataert}, \& {Sharma}}]{pqs09}
{Parrish}, I.~J., {Quataert}, E., \& {Sharma}, P. 2009, \apj, 703, 96

\bibitem[{{Parrish} {et~al.}(2010){Parrish}, {Quataert}, \& {Sharma}}]{pqs10}
---. 2010, \apjl, 712, L194

\bibitem[{{Parrish} \& {Stone}(2005)}]{ps05}
{Parrish}, I.~J., \& {Stone}, J.~M. 2005, \apj, 633, 334

\bibitem[{{Parrish} \& {Stone}(2007)}]{ps07b}
---. 2007, \apj, 664, 135

\bibitem[{{Peterson} \& {Fabian}(2006)}]{pf06}
{Peterson}, J.~R., \& {Fabian}, A.~C. 2006, \physrep, 427, 1

\bibitem[{{Quataert}(2008)}]{quat08}
{Quataert}, E. 2008, \apj, 673, 758

\bibitem[{{Reynolds} {et~al.}(2005){Reynolds}, {McKernan}, {Fabian}, {Stone},
  \& {Vernaleo}}]{reynolds05}
{Reynolds}, C.~S., {McKernan}, B., {Fabian}, A.~C., {Stone}, J.~M., \&
  {Vernaleo}, J.~C. 2005, \mnras, 357, 242

\bibitem[{{Ruszkowski} \& {Oh}(2010)}]{rus10}
{Ruszkowski}, M., \& {Oh}, S.~P. 2010, \apj, 713, 1332

\bibitem[{{Sharma} {et~al.}(2009){Sharma}, {Chandran}, {Quataert}, \&
  {Parrish}}]{sharma09}
{Sharma}, P., {Chandran}, B.~D.~G., {Quataert}, E., \& {Parrish}, I.~J. 2009,
  \apj, 699, 348

\bibitem[{{Sharma} \& {Hammett}(2007)}]{sh07}
{Sharma}, P., \& {Hammett}, G.~W. 2007, Journal of Computational Physics, 227,
  123

\bibitem[{{Spitzer}(1962)}]{spitz62}
{Spitzer}, L. 1962, {Physics of Fully Ionized Gases} (Physics of Fully Ionized
  Gases, New York: Interscience (2nd edition), 1962)

\bibitem[{{Stone} \& {Gardiner}(2007)}]{stoneRT}
{Stone}, J.~M., \& {Gardiner}, T. 2007, \apj, 671, 1726

\bibitem[{{Stone} {et~al.}(2008){Stone}, {Gardiner}, {Teuben}, {Hawley}, \&
  {Simon}}]{sg08}
{Stone}, J.~M., {Gardiner}, T.~A., {Teuben}, P., {Hawley}, J.~F., \& {Simon},
  J.~B. 2008, \apjs, 178, 137

\bibitem[{{Tozzi} \& {Norman}(2001)}]{tn01}
{Tozzi}, P., \& {Norman}, C. 2001, \apj, 546, 63

\bibitem[{{van Leer}(1979)}]{vanLeer}
{van Leer}, B. 1979, Journal of Computational Physics, 32, 101

\bibitem[{{Voit} {et~al.}(2008){Voit}, {Cavagnolo}, {Donahue}, {Rafferty},
  {McNamara}, \& {Nulsen}}]{voit08}
{Voit}, G.~M., {Cavagnolo}, K.~W., {Donahue}, M., {Rafferty}, D.~A.,
  {McNamara}, B.~R., \& {Nulsen}, P.~E.~J. 2008, \apjl, 681, L5

\end{thebibliography}
\end{document}